\documentclass[aps,prl,reprint,superscriptaddress]{revtex4-1}
\usepackage{blindtext}
\usepackage{graphicx}
\usepackage{amsmath}
\usepackage{multirow}
\usepackage{color,soul}
\usepackage[bottom]{footmisc}
\usepackage{url}
\usepackage[colorlinks,
            linkcolor=blue,
            anchorcolor=blue,
            citecolor=blue,
            ]{hyperref}
\usepackage{setspace}

\begin{document}

\title{Efficient Statistical Model for Predicting Electromagnetic Wave Distribution in Coupled Enclosures}

\author{Shukai Ma}
\email{skma@umd.edu}
\affiliation{Quantum Materials Center, Department of Physics, University of Maryland, College Park, Maryland 20742, USA}
\author{Sendy Phang}
\affiliation{School of Mathematical Sciences, University of Nottingham, UK, NG7 2RD}
\affiliation{George Green Institute for Electromagnetics Research, University of Nottingham, UK, NG7 2RD}
\author{Zachary Drikas}
\affiliation{U.S. Naval Research Laboratory, Washington, DC 20375, USA}
\author{Bisrat Addissie}
\affiliation{U.S. Naval Research Laboratory, Washington, DC 20375, USA}
\author{Ronald Hong}
\affiliation{U.S. Naval Research Laboratory, Washington, DC 20375, USA}
\author{Valon Blakaj}
\affiliation{School of Mathematical Sciences, University of Nottingham, UK, NG7 2RD}
\author{Gabriele Gradoni}
\affiliation{School of Mathematical Sciences, University of Nottingham, UK, NG7 2RD}
\affiliation{George Green Institute for Electromagnetics Research, University of Nottingham, UK, NG7 2RD}
\author{Gregor Tanner}
\affiliation{School of Mathematical Sciences, University of Nottingham, UK, NG7 2RD}
\author{Thomas M. Antonsen}
\affiliation{Department of Physics, University of Maryland, College Park, Maryland 20742, USA}
\affiliation{Department of Electrical and Computer Engineering, University of Maryland, College Park, Maryland 20742-3285, USA}
\author{Edward Ott}
\affiliation{Department of Physics, University of Maryland, College Park, Maryland 20742, USA}
\affiliation{Department of Electrical and Computer Engineering, University of Maryland, College Park, Maryland 20742-3285, USA}
\author{Steven M. Anlage}
\affiliation{Quantum Materials Center, Department of Physics, University of Maryland, College Park, Maryland 20742, USA}
\affiliation{Department of Electrical and Computer Engineering, University of Maryland, College Park, Maryland 20742-3285, USA}

\begin{abstract}
The Random Coupling Model (RCM) has been successfully applied to predicting the statistics of currents and voltages at ports in complex electromagnetic (EM) enclosures operating in the short wavelength limit \cite{Hemmady2005,Zheng2006,Zheng2006a,Hemmady2012}.
Recent studies have extended the RCM to systems of multi-mode aperture-coupled enclosures.
However, as the size (as measured in wavelengths) of a coupling aperture grows, the coupling matrix used in the RCM increases as well, and the computation becomes more complex and time consuming.
A simple Power Balance Model (PWB) can provide fast predictions for the \textit{averaged} power density of waves inside electrically-large systems for a wide range of cavity and coupling scenarios.
However, the important interference induced fluctuations of the wave field retained in the RCM are absent in PWB.
Here we aim to combine the best aspects of each model to create a hybrid treatment and study the EM fields in coupled enclosure systems. 
The proposed hybrid approach provides both mean and fluctuation information of the EM fields without the full computational complexity of coupled-cavity RCM.
We compare the hybrid model predictions with experiments on linear cascades of over-moded cavities. We find good agreement over a set of different loss parameters 
and for different coupling strengths between cavities. The range of validity and applicability of the hybrid method  are tested and discussed.

\end{abstract}

\maketitle

\section {I. Introduction}

The ability to characterize and predict the nature of short-wavelength Electromagnetic (EM) waves inside inter-connected enclosures is of interest to various scientific fields. 
Applications include EM compatibility studies for electronic components under high-power microwave exposure \cite{Li2015,Fedeli2009}, coupled quantum mechanical systems modeled with superconducting microwave billiards \cite{Dietz2006}, cascades of quantum dots \cite{Chalker1988,Sau2012, Beenakker2015, Zhang} by way of analogy, and \textit{Smart Homes} sensors in furnished indoor environments.
The enclosures in these applications are generally electrically large with an operating wavelength $\lambda \ll V^{1/3}$, where $V$ is the volume of the system.
The interior geometry of these enclosures is often complex including wall features and internal objects acting as scatterers and geometrical details may not be precisely specified. These systems are then well-described as ray-chaotic enclosures, where the trajectories of rays with slightly different initial conditions diverge exponentially with increasing number of bounces off the irregular walls and interior objects \cite{Dupre2015, Kaina2015, DelHougne2018}. 
This ray-chaotic property has inspired research in diverse contexts such as acoustic \cite{Ellegaard1995, TS07, Auregan2016} and microwave cavities \cite{Doron1990, So1995, Kuhl2005, Gradoni2012, Gagliardi2015, Xiao2018, Ma2019}, the spectral properties of atoms \cite{TRR00} and nuclei \cite{Haq1982}, quantum dot systems \cite{Alhassid2001}. 

Benefiting from the continuing advance in computational capabilities, deterministic approaches utilizing numerical techniques are widely applied in simulating EM quantities inside chaotic systems with specified geometries \cite{Parmantier2004}. The resolution required for deterministic methods, such as the Finite Difference Time Domain (FDTD) or the Finite Element Method (FEM), scale with the inverse of the wavelength and thus consume a large amount of computational resources in the short-wavelength limit (i.e., when typical wavelengths are small compared to the linear scale of the enclosure).
In addition, minute changes of the interior structure of a given system will drastically alter the solution of the EM field \cite{Hemmady2012,Xiao2018}. 
Statistical methods may thus be more appropriate when studying such systems. 
Many such approaches have been proposed, examples include the Baum-Liu-Tesche technique \cite{baum1978analysis, Li2015} which analyzes a complex system by studying the traveling waves between its sub-volumes, and the Power Balance Model (PWB) \cite{Hill1994, Hill1998, Junqua2005, Tait2011a} which predicts the averaged power flow in systems. 
{The PWB method makes predictions of the steady-state averaged energy density inside all system sub-volumes based on equating the incoming and out-going power in each connected sub-volume.}
A version of the PWB method estimating also the variance of the fluctuations has been presented in \cite{Kovalevsky2015}. 
The PWB method is based on the assumption of a uniform field distribution in each cavity which is often fulfilled in the weak inter-cavity coupling and low damping limit. Extensions of the PWB method that drop the uniform field assumption are ray tracing (RT) methods \cite{McKown1991,Savioja2015} or the Dynamical Energy Analysis (DEA) method \cite{Bajars2017, Bajars2017a, HMTC19}  which calculate local ray or energy densities. 
{DEA and RT capture non-uniformity in the field distribution within a given sub-volume. }
Like PWB, they do not treat fluctuations in energy density due to wave interference.

The Random Coupling Model (RCM) \cite{Hemmady2005,Zheng2006,Zheng2006a,Hemmady2012, Zheng2006b,Gradoni2012} allows for the calculation of the statistical properties of EM fields described in terms of scattering and impedance matrices that relate wave amplitudes, or voltages and currents at ports. 
The RCM is based on Random Matrix Theory (RMT) originally introduced to describe complex nuclei \cite{Haq1982}.
{It was later conjectured that any system with chaotic dynamics in the classical limit will also have wave properties whose statistics are governed by random matrix theory} \cite{Casati1980,Bohigas1984} 
{Here the RCM is applied to wave chaotic systems in the short wavelength regime. }
In contrast to the above mentioned methods, the RCM is able to describe the full probability distribution functions (PDFs) for voltages and currents at ports.
{An exemplary application of RCM is the simulation of the fluctuating impedance matrix based on minimal system information, namely a single-cavity loss parameter $\alpha$ (to be defined) and several system-specific features} \cite{Zheng2006,Zheng2006a,Hart2009,Yeh,Gradoni2014,Xiao2016,ma2020}.
{The system-specific features include the radiation information of the ports (both emitting and absorbing), as well as short orbits inside the cavity.
The parameter $\alpha$ reflects the loss level of the system, which can be derived from the overall cavity dimension, Q-factor and operating frequency. }
{The RCM has been successfully applied to single cavities and systems of coupled cavities with varying losses, cavity dimensions, and in the presence of nonlinear elements} \cite{Zhou2017,Zhou2019}.

The computational complexity of the RCM grows, however, with the addition of large apertures connecting together multiple enclosures.
In the RCM an aperture is treated as a set of $M$ correlated ports, the number of which scales with the area of the aperture as measured in wavelengths squared. 
{For example, a circular shaped aperture whose diameter corresponds to four operating wavelengths allows $\sim 100$ propagating modes, leading to $M \sim 100$ ports in the RCM modelling of the inter-connected cavities.}
A cavity with $M$ ports is described by an $(M\times M)$ matrix \cite{Hemmady2005,Zheng2006a,Gradoni2015,ma2020}. 
When large apertures are present, connecting multiple cavities, the RCM model can become cumbersome.
First there is the need to calculate the matrix elements for each aperture that describe the passage of waves through an aperture radiating into free space.
Second these matrices mush be combined with random matrices that give the statistical fluctuations.
Third, the RCM is a Monte-Carlo method in which the matrices simulating the cavities are constructed for each realization, and many realizations may be required to get accurate statistical results.
Finally, the matrices must be connected together which involves inverting the sub-matrices representing the sub-volumes inside a complex system for each realization \cite{ma2020}.
There is thus a need to develop a simple statistical method that applies in cases where the apertures are larger than a wavelength, but small enough so that the two enclosures connected by the aperture can be considered as two separate volumes.

Here, we introduce a hybrid approach that combines the PWB and RCM to generate statistical predictions of the EM field for multi-cavity systems without the computational complexity of a full RCM treatment. The hybrid approach is valid in cases where the coupling between adjacent cavities is carried by many channels due to, for example, large apertures as described above. Using the hybrid method, we apply the PWB method for computing the average EM field intensities in each cavity and use RCM to predict the fluctuations in the cavity of interest only. The modeling of multiple channels between adjacent cavities is thus reduced to computing a scalar coupling coefficient in the PWB model, often given through simple expressions involving the area of the aperture. 
A coupled RCM model still needs to be applied where the number of connecting channels between enclosures is small. Such small apertures act as ``bottlenecks'' for the wave dynamics and a full RCM treatment is necessary to characterize the fluctuations correctly, see also Section V and Appendix C and D. 
From nested reverberation chamber modelling, it is known that weak coupling (bottlenecks) introduces statistical independence of the two cavity environments \cite{Hoijer2013} in the sense that the coupled problem can be described in terms of a random multiplicative process  leading to products of random fields. This leads to strong deviations from a Gaussian random field hypothesis
as discussed in more detail in Section V and Appendix C. 
One may further reduce the computational cost of the hybrid model using a simplified treatment as proposed in Appendix E.
This simplified version of the hybrid model does not require additional knowledge of the frequency-dependent aperture admittance, which is usually obtained through full-wave simulations.

We test the hybrid model using cavity-cascade systems, that is, linear arrays of coupled complex systems. 
RCM based studies for a linear multi-cavity array with a single coupling channel and multiple channels were treated in \cite{Li2015,Gradoni2015, Gradoni2012}. In the following, we introduce the experimental set-up of the cavity cascade system in Section II. The formulation of the PWB-RCM hybrid model is presented in Section III. In Section IV, we compare our model simulations with experimental results. The limits of the hybrid model are studied in Section V with conclusions presented in Section VI. We refer technical details to the Appendices A-E. These contain an in-depth analysis of the limitations of the hybrid model depending on the number of coupling channels in Appendix C, and extension of the hybrid model including ''bottelnecks'' in Appendix D and a simplified version of the original hybrid model in Appendix E.

\section{II. Experimental set-up}

\begin{figure}
\includegraphics[width=0.45\textwidth]{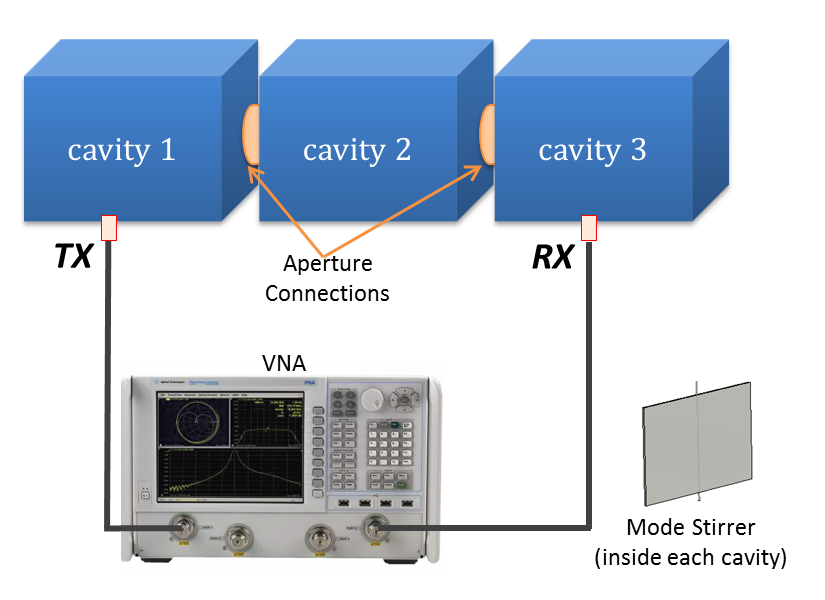}
\caption{\label{fig:3cavpic} A schematic view of the experimental set-up. We measure the $2\times 2$ S-matrix of a cavity cascade system with a VNA between single-mode ports labeled TX and RX. The cavities are connected through multi-mode circular apertures from $1\rightarrow 2\rightarrow 3$. Rotatable mode stirrers are employed in each cavity to generate different system configurations. The 1- and 2-cavity system measurement are conducted by blocking the apertures and moving the location of the RX port to cavity 1 and 2, respectively.}
\end{figure}

We study the transmission and reflection of EM waves in chaotic multi-cavity systems. A series of individual cavities is connected into a linear cascade chain through circular shaped apertures as shown schematically in Fig. \ref{fig:3cavpic}. 
Each cavity is of the same size and shape, but contains a mode-stirrer that makes the wave scattering properties of each cavity uniquely different \cite{Serra2017}. 
The total number of connected cavities is varied from 1 to 3. Short-wavelength EM waves from 3.95 to 5.85 GHz are injected into cavities of dimension $0.762\times0.762\times1.778 m^3$ through WR187 single-mode waveguides, shown as T(R)X in Fig.\ \ref{fig:3cavpic}. 
The loss factor of the system is tuned by placing RF absorber cones inside each cavity.
The cavities are large compared with typical wavelengths of the EM field (with $\lambda =$ 7.49cm - 5.12cm and there are $\sim 10^4$ modes in the frequency range in operation) simulating realistic examples of wave chaotic enclosures. 
The diameter of the aperture is $0.26m$ which requires on the order of $\sim 100$ modes to represent the fields in the aperture at the operating frequency.
The thickness of the aperture is about $0.04$ times the operating wavelength.
We measure the $2\times2$ scattering(S)-matrices of the entire cavity cascade system between ports TX and RX with a Vector Network Analyzer (VNA) from which we deduce the $2\times2$ impedance(Z)-matrix. The S and Z matrices are connected through the bilinear transformation, 
\[
\underline{\underline{S}}=\underline{\underline{Z}}_0^{1/2}(\underline{\underline{Z}}+\underline{\underline{Z}}_0)^{-1}(\underline{\underline{Z}}-\underline{\underline{Z}}_0)\underline{\underline{Z}}_0^{-1/2},\]
where $\underline{\underline{Z}}_0$ is a diagonal matrix whose elements correspond to the characteristic impedances of the waveguide channels leading to the ports. 
Independent mode stirrers are employed inside each cavity to create a large ensemble of statistically distinct realizations of the system \cite{Frazier2013,Frazier2013a,Hemmady,Drikas2014}. 
All mode stirrers are rotated simultaneously to ensure a low correlation between each measurement. A total number of 200 distinct realizations of the cavity cascade are created.

In the RCM, the ``lossyness'' of a single cavity is characterized by the loss parameter $\alpha$ defined as the ratio of the 3-dB bandwidth of a mode resonance to the mean frequency spacing between the modes \cite{Gradoni2014, Xiao2018}. The loss parameter $\alpha$ has values larger than zero (with $\alpha = 0$ corresponds to no loss), where $\alpha$ corresponds roughly to the number of overlapping modes at a given frequency.
For the RCM to be valid, there is an upper limit on $\alpha$ determined by the following conditions: (i) the loss rate and mode density should be relatively uniform over the range of frequencies considered and (ii) the number of overlapping modes should be much less than the number of modes used in the RMT construction of the RCM normalized impedance matrix (see Section III).

The magnitude of the induced voltage at a load attached to the last cavity in the chain, $|U_L|$, can be calculated from the measured impedance matrix $\underline{\underline{Z}}$ of the cavity cascade system \cite{Hemmady2012}. In the experiment, the load is the RX receiver in the VNA ($Z_L = 1/Y_L =  50 \Omega$). Our objective is to use the hybrid PWB-RCM model to describe the statistics of the load voltage $|U_L|$ using a minimum amount of information about the cavities and minimal computational resources.

\section{III. The Hybrid Model}
\subsection{III.A PWB in the Hybrid Model}

The PWB method can be used to determine mean values of EM power flow and energy in systems of coupled cavities \cite{Hill1994,Hill1998,Junqua2005,Tait2011a}. 
For a multi-enclosure problem, the PWB method solves for the mean power density $S_i$ in each enclosure ($i$ is the cavity index) by balancing the powers entering and leaving each cavity.
These power transfer rates are characterized in terms of area cross-sections ($\sigma$), such that the power transferred is $\sigma S_i$.
Various loss channels, such as aperture/port leakage, cavity wall absorption and lossy objects inside the enclosure are characterised through the corresponding cross sections $\sigma_{o}$, $\sigma_{w}$ and $\sigma_{obj}$, respectively \cite{Junqua2005}. 
Constant power is injected into the coupled systems through sources in some or all of the enclosures.
The method solves for a steady state solution when the inputs and losses are made equal for each individual cavity in the system reaching a power balanced state. 
The PWB method does not contain phase information of the EM fields and thus does not describe fluctuations due to interference. 
This can lead to an incomplete prediction of enclosure power flow in the case of small apertures as discussed in Section III.C, Section V.B and Appendix C.

\begin{figure}
\includegraphics[width=0.51\textwidth]{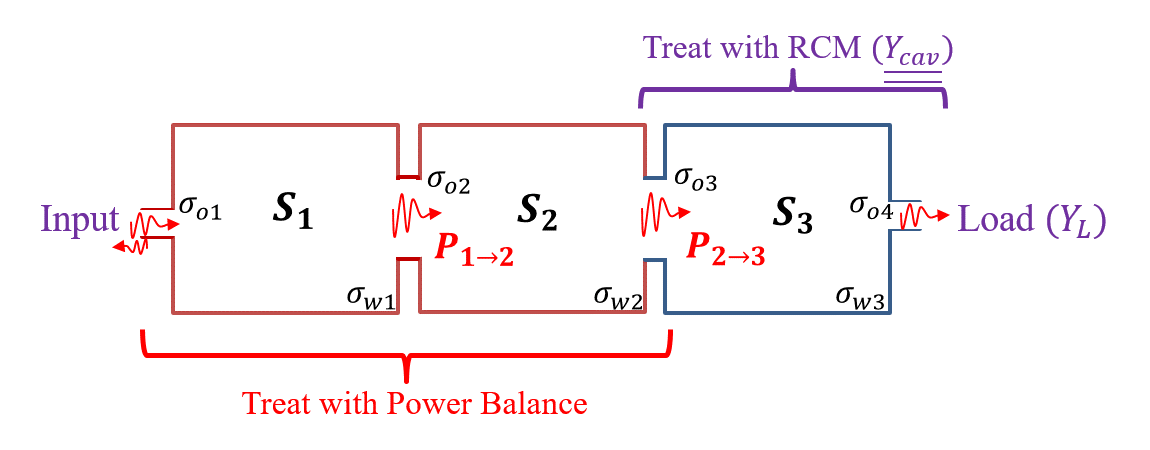}
\caption{\label{fig:hybrid} Schematic illustration of the hybrid model applied to a 3-cavity cascade. In the hybrid model, we use the PWB method to characterize the power flow from the first cavity to the next to last cavity. The fluctuations in the final cavity are described using the RCM method using the mean power flow values obtained from PWB as an input.}
\end{figure}

\subsection{III.B RCM in the Hybrid Model}


{As introduced in Section I, the Random Coupling Model provides an alternative method to describe the statistics of the EM fields in a wide variety of complex systems. In contrast with PWB, the RCM deals with both the mean and fluctuations of the cavity  fields.}

{For coupled-cavity systems, the RCM multi-cavity treatment begins with the modeling of the fluctuating impedance matrix of each individual cavity} \cite{ma2020}.
These matrices relate the voltages and currents at the ports of a cavity. 
When cavities are connected the voltages at the connecting ports are made equal and the connecting currents sum to zero.
The input port on the first cavity is excited with the known signal.
This leads to a linear system of equations that can be solved for all the voltages and currents.
This system is resolved for each realization of the cavity impedance matrices.

{Model realizations of the fluctuating cavity impedance matrix of an individual cavity are created via a normalized impedance matrix $\underline{\underline{\xi}}_{RCM}$ derived from a random matrix ensemble }\cite{Hemmady2005,Hemmady,Zheng2006,Zheng2006a}. 
In terms of the normalized impedance matrix, the fluctuating cavity impedance matrix is written as \[\underline{\underline{Z}}_{cav}=iIm[\underline{\underline{Z}}_{avg}] + Re[\underline{\underline{Z}}_{avg}]^{1/2}\cdot \underline{\underline{\xi}}_{RCM}\cdot  Re[\underline{\underline{Z}}_{avg}]^{1/2},\]
{where the quantity $\underline{\underline{Z}}_{avg}$ is discussed below.}
Here, $\underline{\underline{\xi}}_{RCM}$ is defined as 
\[ \underline{\underline{\xi}}_{RCM}=-\frac{i}{\pi} \sum_n \frac{\underline{w}_{n}\, \underline{w}_{n}^T}{(k_0^2-k_n^2)/\Delta k_n^2+i\alpha},\] 
where the sum over $n$ represents a sum over the modes inside the cavity.
The vector $\underline{w}_{n}$, whose number of elements equals the number of ports, consists of independent, zero mean, unit variance random Gaussian variables which represent the coupling between each port and the $n^{th}$ cavity mode.
This random choice of mode-port coupling originates from the so-called Berry hypothesis, where the cavity modes can be modeled as a superposition of randomly distributed plane waves \cite{Berry1981}.
The quantities $k_0$ and $k_n$ are wavenumbers corresponding to the operating frequency $\omega_0=k_0 \, c$ and the resonant frequencies of the cavity modes, $\omega_n=k_n \, c$.
Rather than use the true resonant frequency of the cavity, a representative set of frequencies is generated from a set of eigenvalues of a random matrix selected from the relevant Random Matrix Theory ensemble \cite{So1995,Fyodorov2004,Hemmady2005,Rehemanjiang2016}.
These RMT eigenvalues are appropriately normalized to give the correct spectral density via the parameter $\Delta k_n^2$.
The loss parameter $\alpha = k^2/(Q\,\Delta k_n^2)$ where $k$ is the wave vector of interest and $Q$ is the quality factor \cite{Gradoni2014}. 

System-specific information about the enclosure is captured in the averaged impedance matrix for each cavity, $Z_{avg}$ \cite{Hart2009}.
Here, the average impedance matrix can be thought of in two ways. 
First it can be considered as a window average of the exact fluctuating cavity matrix over a frequency range, $\omega_2$, centered at frequency $\omega_1$.
In the case in which the windowing function is Lorenzian this average is equivalent to evaluating the exact cavity matrix at complex frequency $\omega_1 +i\, \omega_2$.
This in turn is the response matrix for exponentially growing signals with real frequency $\omega_1$ and growth rate $\omega_2$.
This leads to the second way of understanding the average impedance matrix.
It is the early time ($\omega_2 \cdot t<1$) response of the ports of the cavity.
Thus, the average impedance matrix can be calculated by assuming the walls of the cavity have been moved far from each port, and each port responds as if there were only outgoing waves from the port.
In the case of apertures as ports, the transverse electric and magnetic fields in the aperture opening are expanded in a set of basis modes with amplitudes that are treated as port voltages in the case of electric field, and port currents in the case of magnetic field.
The linear relation between these amplitudes is calculated for the case of radiation into free space, and this becomes the average aperture admittance/impedance matrix.
Each mode in the aperture field representation is treated as a separate port in the cavity matrix.
Thus, the dimensions of the matrix grows rapidly with the addition of a large aperture \cite{Gradoni2015}.

{In the cavity cascade system, the above mentioned apertures (with $M$ propagating modes) are adopted as the connecting channel between neighbouring cavities.
With $M$ connecting channels between cavities, the dimension of the above defined cavity impedance matrix becomes $M\times M$.}
The matrix multiplications and inversions in the calculation of RCM multi-cavity formulations \cite{ma2020} thus have complexity which grows as $O(M^{2.4})$ using common algorithms \cite{Coppersmith1990}. 
Thus for large $M$, the computational cost of the RCM scales roughly as $N\times M^{2.4}$, where N represents the number of cavities in the system.

Here we propose a hybrid method for multi-cavity problems that combines both PWB and RCM as shown schematically in Fig. \ref{fig:hybrid}. In an $N-$cavity cascade system with multiple channel connections between adjacent cavities, we utilize the PWB to characterize the mean flow of EM waves from the input port of the first cavity to the input aperture of the last ($N^{th}$) cavity. 
The RCM method is now applied to the last cavity and the connected load at the single-mode output port of that cavity. 
Thus the hybrid method combines the strengths of both methods: the fluctuations of the EM field will be captured with RCM, and the computational cost is greatly reduced using the PWB method.
{A quantitative comparison of the computational costs between full RCM method and the hybrid method is discussed later in Section III.D.}

In the following, we discuss the hybrid model formulation in detail based on the 3-cavity system example shown in Fig. \ref{fig:hybrid}.
We will first introduce the PWB treatment to the first two cavities in the chain, followed by the modeling of the last cavity using RCM in subsection III.C. 
We will then discuss how to connect the PWB and RCM models at the aperture plane between the last two cavities in subsection III.D. 
With the model formulation introduced, we will look into the validity of the hybrid model in subsection III.E.
A step-by-step protocol to apply the PWB-RCM fusion to generic cavity systems is detailed in Appendix D.

\subsection{III.C Detailed Cavity Treatments}

PWB characterizes the flow of high-frequency EM waves inside a complex inter-connected system based on the physical dimensions of the cavities, the cavity quality factors $Q$, and the coupling cross sections $\sigma_{o}$ as well as the incident power $P_{in}$ driving the system \cite{Hill1994,Junqua2005}. PWB then calculates the power densities of each individual cavity in steady state. For the 3-cavity cascade system in Fig. \ref{fig:hybrid}, the PWB equations are
\begin{equation} \label{eq:pwb}
\resizebox{0.5\textwidth}{!}
    {$\left[ \begin{matrix} \sigma_{w1}+\sigma_{o1}+\sigma_{o2} & -\sigma_{o2} & 0 \\
    -\sigma_{o2} & \sigma_{w2}+\sigma_{o2}+\sigma_{o3} & -\sigma_{o3} \\
    0 & -\sigma_{o3} & \sigma_{w3}+\sigma_{o3}+\sigma_{o4}\\
    \end{matrix} \right] \left[ \begin{matrix} S_{1} \\ S_{2} \\ S_{3} \end{matrix} \right] = \left[ \begin{matrix} P_{in} \\ 0 \\ 0 \end{matrix} \right]$}
\end{equation}
where the $\sigma_{wi}$'s  refer to the wall loss cross section and $S_{i}$ is the power density of the $i^{th}$ cavity, $\sigma_{o1}$ and $\sigma_{o4}$ are the cross section of the input and output ports, while $\sigma_{o2}$ and $\sigma_{o3}$ represent the aperture cross sections, see Fig. \ref{fig:hybrid}. 
The cross sections can be expressed explicitly from known physical dimensions and cavity wall properties \cite{Junqua2005}. $P_{in}$ is the (assumed steady) incident power flow into the first cavity. In this example, we assume that only the first cavity receives EM power from external sources. 
The balance between the input and loss is achieved by solving Eq.\ (\ref{eq:pwb}) for the steady state power densities $S_{i}$. For example, the power balance condition of the first enclosure is expressed as $(\sigma_{w1}+\sigma_{o1}+\sigma_{o2})\cdot S_{1}=P_{in}+\sigma_{o2}\cdot S_{2}$. The LHS of this equation represent the loss channels of the cavity, including the cavity wall loss and the leakage through the input port and the aperture. The RHS describes the power fed into the cavity, consisting of the external incident power and the power flow from the second cavity. The net power that flows into the last cavity in the cascade is expressed as $P_{2\rightarrow 3}=\sigma_{o3}(S_{2}-S_{3})$.

The last cavity is characterized by the RCM method. With the knowledge of the cavity loss parameter $\alpha$ and the port coupling details, the full cavity admittance matrix of the last cavity can be expressed as \cite{Gradoni2015}
\[
\underline{\underline{Y}}_{cav}=i\cdot Im\left(\underline{\underline{Y}}_{rad}\right) \, + \, Re\left(\underline{\underline{Y}}_{rad}\right)^{0.5}\cdot \underline{\underline{\xi}}\cdot Re\left(\underline{\underline{Y}}_{rad}\right)^{0.5}.
\]  
The quantity $\underline{\underline{Y}}_{rad}$ is a frequency-dependent block-diagonal matrix whose components are the radiation admittance matrices of all the ports and apertures of that cavity. 
We assume no direct couplings between apertures because the direct line-of-sight effect is small in the experimental set-up.
Consider a cavity with two $M-$mode aperture connections, the dimension of the corresponding matrix $\underline{\underline{Y}}_{rad}$ is $2M\times 2M$. 
The matrix elements are complex functions of frequency in general and can be calculated using numerical simulation tools. 
We use here the software package {\em CST Studio} to calculate the aperture radiation admittance (see Ref. \cite{Gunnarsson2014} and the Appendix B.2 in Ref. \cite{ma2020}).
The RCM normalized impedance $\underline{\underline{\xi}}$ is a detail-independent fluctuating ``kernel'' of the total cavity admittance $\underline{\underline{Y}}_{cav}$. With known $\alpha$, an ensemble of the normalized admittance $\underline{\underline{\xi}}$ can be generated through random matrix Monte Carlo approaches \cite{Hemmady2012}. Combining the fluctuating $\underline{\underline{\xi}}$ and $\underline{\underline{Y}}_{rad}$, an ensemble of ``dressed'' single cavity admittance matrices for the final cavity can be generated.
It is later shown in Appendix E that a substantial reduction of the hybrid model computational cost is made possible using an aperture-admittance-free treatment, at the price of reduced accuracy for longer cascade chains.

\subsection{III.D The Hybrid Model}

We next connect the PWB and RCM treatments at the interface between the second and the third cavity. As discussed in the previous section, the power flow into the third cavity, $P_{2\rightarrow 3}$, is calculated from the 3-cavity PWB calculation. Identical system set-ups are utilized in the PWB and RCM treatments, including the operating frequency range, the dimensions of the cavities, ports and apertures, and the loss of the cavity (achieved by a simple analytical relationship between the RCM $\alpha$ parameter and $\sigma_{w}$, see Appendix A for more details).
To transfer the scalar power values $P_{2\rightarrow 3}$ generated by PWB into an aperture voltage vector required for RCM, we assign random voltages $\underline{U}_{o3}$ drawn from a zero mean, unit variance Gaussian distribution for the $M$-mode aperture and calculate the random aperture power using 
\[P_{o3}=\frac{1}{2}Re(\underline{U}_{o3}^*\cdot \underline{\underline{Y}}_{rad} \cdot \underline{U}_{o3}).\] 
These randomly assigned aperture voltages $\underline{U}_{o3}$ are then normalized by the ratio $P_{2\rightarrow 3}/P_{o3}$ to match with the value calculated from PWB. Combined with the RCM generated cavity admittance matrix $\underline{\underline{Y}}_{cav}$, an ensemble of induced voltage values $U_L$ at the load on the last cavity is computed utilizing Eq. (A.9) in Ref. \cite{ma2020}. The power delivered to the load is obtained using 
\[P_L=\frac{1}{2}Re({U_{L}}^*\cdot {Y_{L}} \cdot {U_{L}}),\] 
where $Y_L$ is the load admittance, taken to be $1/(50\Omega)$ here. 

{With the formulation of the hybrid model now explained, here we discuss the improvement in computational time and memory usage by replacing the full RCM multi-cavity method with the hybrid model.
For an $N-$cavity system connected through $M-$mode apertures, the hybrid model requires only a fraction of $1/[(N-1)M^2]$ the memory consumption as compared to the full RCM method, enabled by the reduced cavity impedance matrix storage for the first $N-1$ cavities.}
{Moreover, the speed of computation is also reduced by eliminating the RCM modeling of the prior cavities.}
{For example, the computation time for the two-cavity cascade system with circular aperture connections reduces from about 120s to 40s with application of the hybrid method (tests run on a typical workstation). 
In addition, the computation time of the full RCM method scales with the total number of cavities in the system N, while the hybrid method is insensitive to the further addition of cavities.} 
{The advantage of the method becomes more prominent for larger lengths of the cavity chain $N$ and larger apertures (having M modes).}

\subsection{III.E Limits of the Hybrid Model} \label{sec:limits}

The hybrid model is based on the assumption that the fluctuations in a given cavity are independent of the fluctuations in adjacent cavities and thus of the fluctuations in the power flowing between cavities (as a function of frequency, for example). We assess the validity of these assumptions by analysing a multi-channel cascaded cavity system in Appendix C. We study in particular the effect of the total number of effective cavity-cavity coupling channels $M_n$ between the $n$th and $(n+1)$st cavity on the fluctuation levels of the load-induced power $P_L$ connected to the final cavity. Since $P_L \propto |U_L|^2$, the conclusions drawn from the power flow studies can also be applied to voltage-related results as presented below in Section IV.
Defining the load power fluctuation levels as the ratio $\kappa=\left< P_L \right>^2/\left< P_L^2 \right>$, where $\left< \cdots \right>$ represents averaging over an ensemble, we treat $\kappa$ as a measure characterising the level of fluctuations of the power. Here, $\kappa\sim 1$ and $\kappa \gg 1$ refer to low and high fluctuations of the power values, respectively. 
In Appendix C, it is shown that 
\begin{equation}\label{prod} \kappa \propto \prod_{n=1}^{N}(1+M_n^{-1}),\end{equation}
where the product is over all the cavities in the cascade. 
If cavities $n=1$ to $n = N-1$ have strong multi-channel connections due to, for example, large apertures with $M_n\gg 1$ propagating modes at the operating frequency, then $M_n^{-1} \rightarrow 0$ and the contributions to (\ref{prod}) not including the coupling to the load, that is, $\prod_{n=1}^{N-1}(1+M_n^{-1}) \rightarrow 1$; this holds for the experiments described in Section IV with $M_n \approx 100$ at the frequencies considered. 
The quantity $M_n$ is small when a single-mode waveguide connects the last cavity ($N$) to the load (the experiment in Section IV).
At the last cavity ($N$) there is a single mode output port, $M_n=1$ and the $1+M_n^{-1} = 2$ which induces higher fluctuations at the load compared to the case where all apertures are large. 
Similar small $M_n$ situations appear when a ``bottleneck'' is introduces between cavities (the experiments in Section V.B).
It is therefore sensible to adopt RCM for just the last cavity to capture the power fluctuations at the load, while it is sufficient to include the influence of the intervening cavities with PWB only giving the required information about the mean power flow. This case is discussed in Section IV. In Section V, we will also consider the effect of having small aperture connections -- referred to as ``bottlenecks'' -- at intermediate locations in the cavity cascade where we see deviations of the hybrid model from a full multi-cavity RCM treatment.

\section{IV. Comparison of Hybrid Model with Experimental Results and Discussion}

\begin{figure}
\centering
\includegraphics[width=0.5\textwidth]{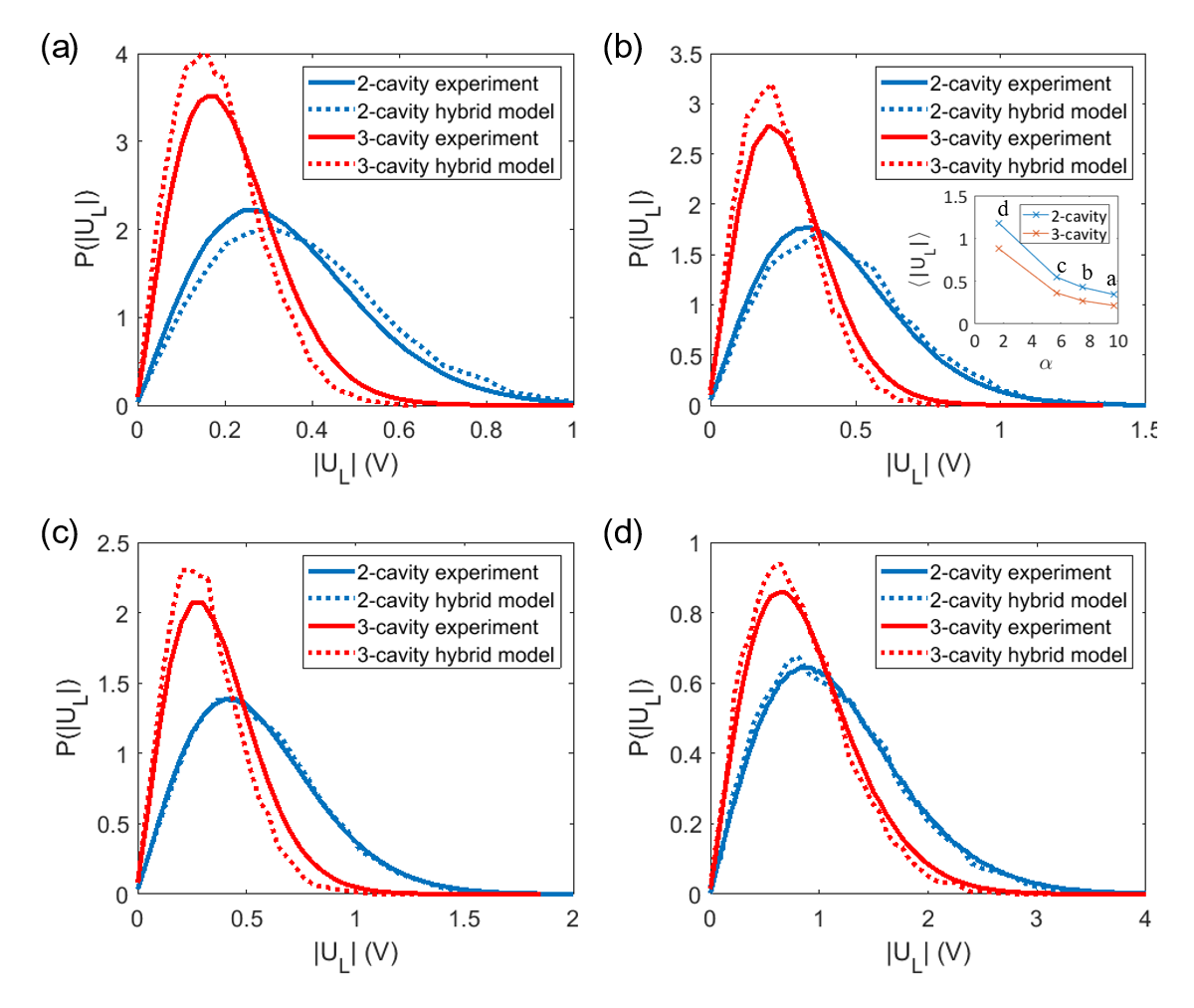}
\caption{\label{fig:induceV_a} The PDFs of load induced voltage $|U_L|$ of 2- and 3-cavity experiments (solid) and hybrid model calculations (dashed). The single cavity loss parameter is varied from 9.7, 7.5, 5.7 and 1.7 from (a-d), respectively. The inset in (b) shows the 2- and 3-cavity experiment averaged induced voltage values $\langle |U_L| \rangle$ with respect to different loss parameters $\alpha$. Multi-mode ($\sim 100$ modes) circular apertures are employed between the cavities.}
\end{figure}

We now conduct the PWB-RCM hybrid analysis for  the multi-cavity experiment and compare with measurements. 
We consider 2- and 3-cavity cascades with large (on the scale of the wavelength) apertures between the cavities and single-mode connections to the load in the last cavity.
The induced voltages at the load $|U_L|$ are calculated from data using the methods reported in Refs. \cite{Hemmady2012,GilGil2016}, and these experimental results are shown as solid lines in Fig.\ \ref{fig:induceV_a}.  
An ensemble of induced load voltages for the multi-cavity system is created by moving the mode stirrers in all cavities between each measurement. 
The losses in the single cavities is altered by inserting equal amounts of RF absorbers in each cavity.
In addition, the hybrid PWB-RCM method is used to calculate $|U_L|$ and the resulting distributions are shown as dotted lines in Fig. \ref{fig:induceV_a}.
Good agreement between the measured and model generated results are observed over a range of different total cavity numbers and single cavity loss values. 
Under varying cavity loss conditions, the probability density function (PDF) of the induced load voltage $|U_L|$ of the 3-cavity system has a lower mean value and smaller fluctuations compared to the 2-cavity system results. Going from two to three cavities will decrease the energy density in the last cavity and thus the power delivered to the load.
This difference between the 2- and 3-cavity $|U_L|$ becomes smaller when the single cavity loss decrease as can be seen following Fig. \ref{fig:induceV_a} (a) to Fig. \ref{fig:induceV_a} (d), see also the inset in Fig. \ref{fig:induceV_a} (b). 

\begin{figure}
\includegraphics[width=0.45\textwidth]{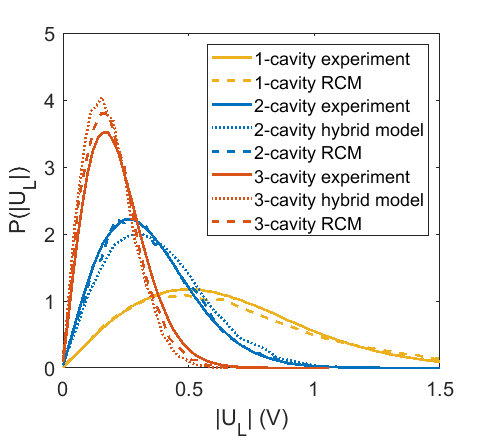}
\caption{\label{fig:induceV_rcm} Comparison of induced load voltage statistics for the hybrid model (dotted), RCM predictions (dashed) and experimental results (solid) for the case of 2 and 3 cavity cascades with single-cavity loss parameter $\alpha=9.7$, and circular multi-mode apertures between the cavities. The frequency range is from 3.95 to 5.85 GHz.}
\end{figure}

The induced load voltage PDFs of the multi-cavity system can be generated solely with the RCM formulation \cite{ma2020}. A comparison between the $|U_L|$ PDFs generated with full RCM method, the hybrid method, and the experiments are shown in Fig.\ \ref{fig:induceV_rcm}. 
Both theoretical approaches are able to generate statistical ensembles which agree well with the experimental results. 
We find that the RCM results (dashed lines) slightly outperform the hybrid method (dotted lines) for the two-cavity case. However, the computation time and storage cost of the full RCM method is $N$ times larger than the hybrid method, where $N$ refers to the total number of connected cavities in the cascade.

{It is well-established that the distribution of fields inside sufficiently lossy single/nested cavity systems is Rayleigh/double-Rayleigh distributed} \cite{Zhou2017, Hoijer2013, Yeh2012, Yeh2013}.
{It was shown by the authors of Ref.} \cite{Zhou2017} {that the cavity field distribution deviates from a Rayleigh distribution in the low-loss limit ($\alpha<1$), while the RCM method remains valid.
Thus the RCM method applies to a wider range of complex cavity systems. }

\section{V. Testing the limits of the hybrid model} \label{sec:limits_detailed}

As already discussed in Section III C, we expect that the hybrid model will break down in the limit where either the RCM and/or PWB methods are no longer valid. We consider two generic types of limitations, namely systems with high loss and systems having ``bottlenecks'' in the middle of the cavity cascade. These two conditions are experimentally studied using a $\times 20$ scaled down version of the cavity system \cite{Xiao2018,ma2020}. The dimension of the single miniature cavity is $0.038\times 0.038 \times 0.089m^3$. EM waves from 75-110GHz are fed into the cavity ($\sim 10^4$ modes at and below the operating frequency range) through single-mode WR10 waveguides from Virginia Diodes VNA Extenders and the S-parameters are measured by a VNA. We have previously demonstrated that identical statistical electromagnetic properties are found in this and the full-scale configuration described in Section II \cite{ma2020}. 

\begin{figure}
\includegraphics[width=0.5\textwidth]{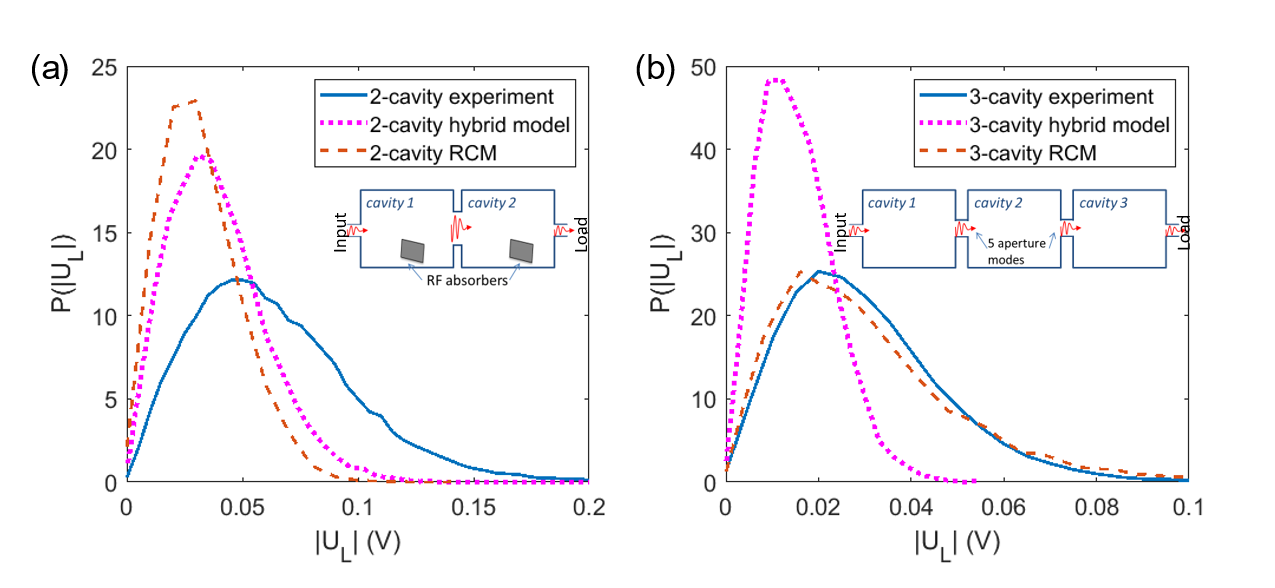}
\caption{\label{fig:breakdown} (a) The load induced voltage $|U_L|$ statistics $P(|U_L|)$ of the 2-cavity experiment and models in the high loss limit $(\alpha \sim 25)$. The cavities are connected with a circular shaped aperture which supports $\sim 100$ modes at 110GHz. The inset shows the experimental set-up schematically. (b) 3-cavity experimental and model generated load induced voltage statistics. The cavities $(\alpha \sim 9.1)$ are connected by rectangular shaped apertures with just 5 propagating modes. The inset is the schematic of the experimental set-up.}
\end{figure}

\subsection{V.A High and Inhomogeneous Cavity Losses}

The proposed hybrid method is not expected to generate accurate predictions for extreme high and inhomogeneous lossy systems \cite{GilGil2016}. Both PWB and RCM models require uniform power distribution inside the studied system. Such a presumption no longer holds when the loss of the cavity wall becomes so high such that the energy distribution near the system boundaries and from the input to the output aperture drop considerably. In this case, PWB needs to be replaced with other methods such as ray tracing \cite{Savioja2015} or the DEA analysis \cite{HMTC19} or, in the case of multiple scatterers in each cavity, using an approximate flow solver based on a 3D diffusion model \cite{Yan2019}; all these methods have a  larger computational overhead compared to PWB. Strong damping also violates the random plane wave hypothesis crucial to the RCM \cite{GilGil2016, Berry1981, Wu1998}. We experimentally examine the applicability of the hybrid model in the extremely high loss limit. 
A 2-cavity cascade system is designed where electrically-large (many wavelength in size) ARC RF absorber panels are placed on a wall inside each cavity. The detailed experimental set-up of the extreme high-loss cases can be found in Appendix B. The inclusion of absorber walls creates effectively ``open-wall'' high loss cavities ($\alpha \sim 25$ for a single cavity) \cite{GilGil2016}. 
The measured and the model-generated load induced voltage statistics are shown in Fig. \ref{fig:breakdown} (a) for this case. Neither the hybrid model nor the full RCM model is expected to work in this high and inhomogenous loss limit. As seen in Fig. \ref{fig:breakdown} (a), both models show strong disagreement with the measured data. These inhomogeneities in each cavity are well captured using either ray-tracing or DEA methods as demonstrated in \cite{Adnan2020}. 

The hybrid model can be applied to the lower loss systems ($\alpha \sim 1$) as demonstrated in Section IV. We point out, however, that the stronger impedance fluctuations of low-loss systems poses greater challenges for the acquisition of good statistical ensemble data for both numerical and experimental methods \cite{Yeh2013,Yeh2013a}.

\subsection{V.B Weak Inter-Cavity Coupling}

Another assumption of the hybrid model is that intermediate cavities have multiple connecting channels, see Section III C. We expect that the hybrid model as introduced in Section III fails when some or all of the apertures are small, effectively acting as ``bottlenecks''. 
The transmission rates then become strongly frequency dependent thus adding to the overall fluctuations in the system and deviating strongly from the aperture cross sections assumed for the PWB in Section III.E. In the experiments, we create a ``bottleneck'' carrying only 5 channels between the intermediate cavities of the chain and study how this brings out the limitations of the hybrid model. 
As discussed in Section III.E and Appendix C, the fluctuations of the wave flow are then correlated as the energy propagates through the cavity cascade chain; these increased fluctuations of the input power at the last cavity are not captured in a PWB treatment.
It is important to point out that the fluctuations of input current at a cavity beyond a bottleneck can no longer be considered Gaussian. The latter aspect is known from cavity systems studied in electromagnetic compatibility, see for example the deviation of the received power in a weakly coupled nested reverberation chamber \cite{Hoijer2013}. 
In Appendix C, arguments are developed to quantify these deviations for cavity cascade systems.

We construct a 3-cavity experiment to test this effect; for more details about the experiment, see Appendix B. The cavities  are connected through small rectangular shaped apertures which allow only 5 propagating modes as opposed to 100 modes utilized in experiments discussed earlier; we have $\alpha=9.1$ for all cavities. 
In this few-coupling-channel system, the cavities in the cascade have a substantial contribution to the overall fluctuations of neighbouring cavities and thus the induced load 
voltage/power values. As shown in Fig. \ref{fig:breakdown} (b), the hybrid model fails to correctly reproduce the statistics of the measured $|U_L|$, while the full RCM cavity cascade formulation retains the ability to correctly characterize the statistics of the system.

To summarize the section, we study the applicability of the PWB-RCM hybrid model to cases where it is expected to fail.
It is found that enclosures with large and inhomogeneous losses violate the conditions for the hybrid method.
The case of intermediate ``bottlenecks'' can be handled with RCM, but not the hybrid model as formulated.
A generalization of the PWB-RCM hybrid that can handle arbitrary ``bottlenecks'' is outlined in Appendix D.
The proposed hybrid model is expected to generate statistical mean and fluctuations of the EM fields for systems with moderate cavity losses and large inter-cavity couplings strengths.  Expanding the hybrid model to systems with more complex topology and exploring new types of hybrid models by combining RCM with other statistical or deterministic methods has been considered in \cite{Lin2019}.

\section{VI. Conclusion}

In this manuscript, we propose a PWB-RCM hybrid model for the statistical analysis of EM-fields in complex, coupled cavity systems based on minimal information of the system. The method is tested and found to be in good agreement with cavity cascade experiments under various conditions, such as varying single cavity loss and the total number of connected cavities. The limitations of the hybrid model are also discussed and demonstrated experimentally. The hybrid model is computationally low-cost and able to describe the statistical fluctuations of the EM fields under appropriate conditions. We believe that the hybrid method may find broad applications in the analysis of coupled electrically large systems with sophisticated connection scenarios.

\begin{acknowledgments}
We thank B. Xiao who designed the scaled cavity systems and M. Zhou for the help in conducting numerical computer simulations. We are thankful for R. Gunnarsson's warmhearted help with computer simulation techniques.
This work was supported by the Office of Naval Research under ONR Grant No. N000141512134, ONR Grant No. N000141912481, ONR DURIP Grant N000141410772 and ONR Grant No. N62909-16-1-2115 as well as by the the Air Force Office of Scientific Research  AFOSR COE Grant FA9550-15-1-0171.
\end{acknowledgments}

\appendix*
\setcounter{figure}{0} \renewcommand{\thefigure}{A.\arabic{figure}}

\begin{figure}
\includegraphics[width=0.48\textwidth]{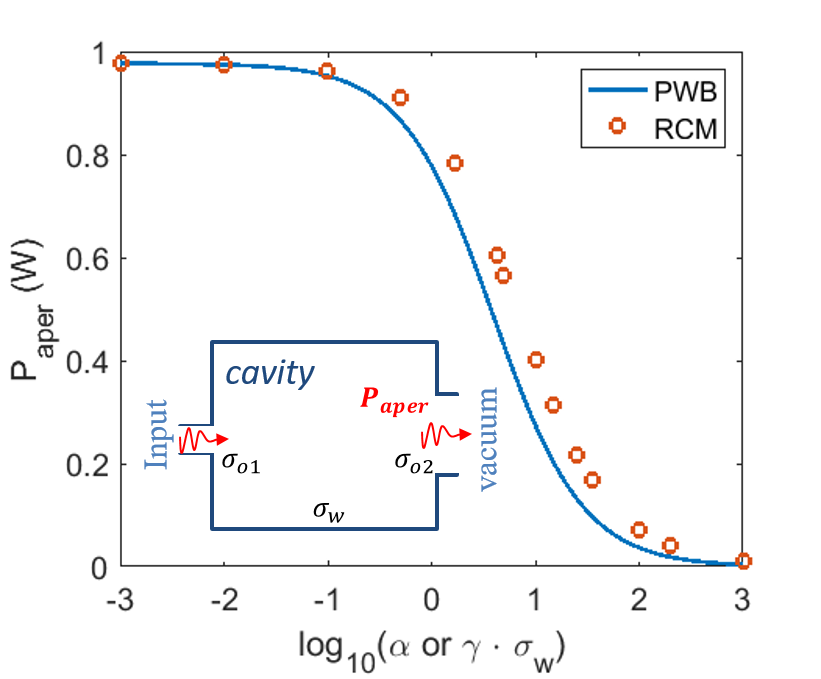}
\caption{\label{fig:singcav} The single-cavity radiated power through the aperture opening $P_{aper}$ vs the cavity loss calculated with PWB $(\gamma \cdot \sigma_w)$ and RCM methods $(\alpha)$. Input power is set as 1 Watt. Inset: schematic diagram of the single cavity radiation set-up.}
\end{figure}

\section{Appendix A: Relating Cavity Internal Loss Parameters in RCM and PWB}

The parameters used to model the cavity loss in the RCM ($\alpha$) and PWB ($\sigma_w$) are different.  We propose a simple analytical relationship between these internal loss parameters, namely $\alpha=\gamma \cdot \sigma_w$. 
In the RCM, the single cavity loss parameter $\alpha$ can be written as $\alpha = k^3 V/(2\pi^2 Q)$ for three-dimensional enclosures, where $V$, $Q$ and $k$ are the cavity volume, closed cavity quality factor and the operating wavenumber \cite{Xiao2018,Gradoni2014}. 
In the PWB, the wall loss cross-section $\sigma_w$ is defined as $\sigma_w = P_w / S$ where $P_w$ is the steady state power loss caused by wall absorption, and $S$ represents the unidirectional power density at any point inside the cavity.
To arrive at an analytical expression for $S$, we first write the Poynting vector $S_p$ which flows uniformly in all directions as $S_p = cW/(4\pi \cdot V)$, where $W$ is the stored energy inside the cavity volume $V$. 
Thus $S$ can be calculated through the integral
\[
S = \int_0^{2\pi} d\phi \int_0^{\frac{\pi}{2}} cos\theta \, S_p \, sin\theta \,d\theta = \frac{cW}{4V}.
\]
Combined with $Q=\omega W/P_w$ (which assumes that wall loss dominates the closed-cavity Q), we have
\[
\sigma_w = \frac{P_w}{S} = \frac{\omega W}{Q} \cdot \frac{4V}{cW} = \frac{4kV}{Q}
\]
and finally arrive at $\alpha = \gamma \cdot \sigma_w$ where $\gamma = 1/(2\lambda^2)$.

We next conduct a sanity check of the $\alpha$ to $\sigma_w$ relationship by creating a practical scenario.
Consider a single lossy cavity with a single input port and a radiating aperture.
We choose this particular scenario because it reflects the range of situations where we believe the RCM/PWB hybrid model will be relevant and valid.
We calculate the radiated power from the cavity as a function of the internal loss parameter in each model, for fixed input power.
In the PWB treatment, a single-mode port and an aperture are opened on the cavity with their PWB cross sections $\sigma_{o1}$ and $\sigma_{o2}$, respectively. Waves are incident into the cavity through the port and exit the cavity through either the aperture or the port into vacuum. In the RCM treatment of the same system, the port and the aperture are modelled with their corresponding radiation impedance (for the port), and radiation admittance (for the aperture).

For the study shown in Fig.\ \ref{fig:singcav}, the dimension of the single cavity is set to $0.762\times0.762\times1.778 m^3$, which is the exact cavity dimension used in the experimental set-up.
The aperture radiated power $P_{aper}$ is calculated with both treatments at 5GHz under various cavity loss values. The aperture is set to be a circular shaped aperture which allows $\sim 75$ propagating modes at 5GHz. 
The results for RCM are shown as red circles in Fig.\ \ref{fig:singcav}. The PWB results are obtained for a series of $\sigma_w$ values and shown as a blue line in Fig. \ref{fig:singcav}.
The factor $\gamma = 1/(2\lambda^2) = 139 m^{-2}$ is applied for the PWB calculation to scale the $\sigma_w$ value to $\alpha$ on the horizontal axis.
A good agreement of aperture radiated power is found between the two models.
We note that a finite discrepancy is observed in the range $\alpha \in [0.1, 100]$.
This slight difference may be caused by the presence of an aperture whose dimension is comparable to the operating wavelength (the resonance regime).
In this limit, the PWB formulas developed for electrically small/large apertures may require modification.

\section{Appendix B: The Experimental Set-up of Hybrid Model Limitation Studies}

\begin{figure}
\includegraphics[width=0.5\textwidth]{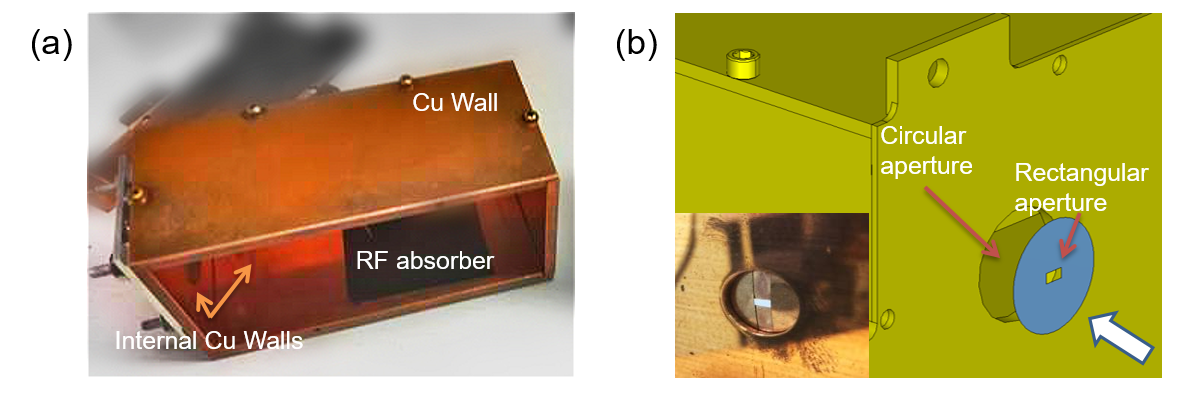}
\caption{\label{fig:hybridlimits} (a) The open-wall view of the single cavity under extreme high-loss conditions. A piece of the RF absorber material (black rectangular shaped) is attached at one cavity wall. (b) The schematic picture of how the rectangular shaped aperture is created. Copper tape with a small rectangular opening (shown in blue) is placed over the original circular aperture. The inset shows a picture of the rectangular shaped aperture.}
\end{figure}

As discussed in section III and IV, both the PWB and RCM methods assume that the wave energy inside the sub-volumes are homogeneously distributed. We would not expect the hybrid model to be effective when this assumption is violated. We experimentally test this limit of the hybrid method by creating a system with non-uniform energy distributions (Fig.\ \ref{fig:hybridlimits} (a)). In the 2-cavity cascade set-up, we first cover one of the cavity walls with RF absorbing materials to create an effective `open-window' cavity. 
The results for the load induced voltage statistics in this cavity can be found in Fig.\ \ref{fig:breakdown} (a). The single cavity loss parameter is $\alpha=25$, estimated by  calculating the Q-factor of the S-parameter measurements \cite{Xiao2018}. It is shown that neither the hybrid method nor the RCM method can correctly reproduce  the experimental results for the load induced voltage statistics.

We also test the case where the apertures are changed from a large circular shaped aperture with approximately 100 propagating modes into a small rectangular shaped aperture with only 5 propagating modes (Fig.\ \ref{fig:hybridlimits} (b)). As shown in Fig.\ \ref{fig:breakdown} (b), the hybrid model prediction deviates from the experimental results, while the full RCM prediction retains the ability to accurately predict the induced-voltage PDF. 

\section{Appendix C: Multi-channel analysis for the cavity cascade system}

Within the RCM description, the $n^{th}$ cavity in the cavity cascade chain can be modeled by the following equation:
\begin{equation} \label{eq:cav}
    \left[ \begin{matrix} \underline{V}_n^i \\ \underline{V}_n^o \end{matrix}\right]=\left[ \begin{matrix}  \underline{\underline{Z}}_n^{ii} & \underline{\underline{Z}}_n^{io} \\ \underline{\underline{Z}}_n^{oi} & \underline{\underline{Z}}_n^{oo}  \end{matrix}\right]\, \left[ \begin{matrix} \underline{I}_n^i \\ \underline{I}_n^o \end{matrix}\right].
\end{equation}
The superscripts $i$ and $o$ refer to the input and output side of the $n^{th}$ cavity where we adopt the convention that ``input" is on the side closest to the input TX port of the cavity chain and ``output" is on the side with the RX port (see Fig. \ref{fig:3cavpic}). The voltage vectors $\underline{V}$ and current vectors $\underline{I}$ at the input and output sides of a cavity are connected through the cavity impedance matrix as shown in Eq.\ ((\ref{eq:cav})). 
The Z-matrix of the $n^{th}$  cavity is written as a $2\times 2$ block matrix. 
Assuming the $n^{th}$ cavity has $m_1$ and $m_2$ coupling channels at its input and output sides, the input vectors ($\underline{V}_n^i$, $\underline{I}_n^i$) are $m_1 \times 1$, and the output vectors ($\underline{V}_n^o$, $\underline{I}_n^o$) are $m_2 \times 1$. Correspondingly, the dimension of the Z-matrix elements are $m_1\times m_1$, $m_2\times m_2$, $m_1\times m_2$ and $m_2\times m_1$ for $\underline{\underline{Z}}_n^{ii}$, $\underline{\underline{Z}}_n^{oo}$, $\underline{\underline{Z}}_n^{io}$ and $\underline{\underline{Z}}_n^{oi}$, respectively.
For simplicity, we assume the RCM description of the off-diagonal impedance matrices as 
\begin{equation} \label{imp_off}
\underline{\underline{Z}}_{n}^{oi}=(\underline{\underline{R}}_{n}^{o})^{1/2}\underline{\underline{\xi}}_n(\underline{\underline{R}}_{n}^{i})^{1/2},
\end{equation}
 and the matrices $\underline{\underline{Z}}_{n}^{oo,ii}$ are diagonal. The $\underline{\underline{R}}_{n}^{i,o}$ are the input and output radiation resistance matrices, and $\underline{\underline{\xi}}_n$ is the RCM normalized impedance matrix.
Since the output side of the $n^{th}$ cavity is connected to the input side of the $(n+1)^{st}$ cavity, the continuity relationships of voltages and currents between the two cavities are:
\begin{equation} \label{eq:cont}
    \underline{V}_n^o=\underline{V}_{n+1}^i, \quad \underline{I}_n^o=-\underline{I}_{n+1}^i.
\end{equation}
In the high-loss limit, the output current at the $(n+1)^{st}$ cavity $\underline{I}_{n+1}^o$ is much smaller than the input current at the $n^{th}$ cavity $\underline{I}_{n}^i$.
By solving Eqs.\ (\ref{eq:cav}) and (\ref{eq:cont}) and neglecting terms $\underline{\underline{Z}}_{n+1}^{io}\, \underline{I}_{n+1}^o$ (small compared to the $\underline{\underline{Z}}^{ii}$ terms), the relationship between the input currents at the $n^{th}$ and $(n+1)^{st}$ cavity in the high-loss limit is written as  
\[(\underline{\underline{Z}}_{n+1}^{ii}+\underline{\underline{Z}}_n^{oo})^{-1}\underline{\underline{Z}}_n^{oi}\,\underline{I}_n^i=\underline{I}_{n+1}^i.
\] 
We then rewrite this equation as:
\begin{equation} \label{eq:a6}
\begin{aligned}
   \underline{\psi}_{n+1} &=\underline{\underline{\tau}}_{n\rightarrow n+1}\,\underline{\underline{\xi}}_n\,{\underline{\psi}_n}
\quad \mbox{with} \\
\quad \underline{\underline{\tau}}_{n\rightarrow n+1}&=(\underline{\underline{R}}_{n+1}^{i})^{1/2}(\underline{\underline{Z}}_{n+1}^{ii}+\underline{\underline{Z}}_n^{oo})^{-1} (\underline{\underline{R}}_{n}^{o})^{1/2},
\end{aligned}
\end{equation}
a transition matrix which describes the coupling from the $n^{th}$ cavity to the $(n+1)^{st}$ with $\underline{\underline{R}}_{n}^{i,o}$ as in (\ref{imp_off}), and the $\underline{\psi}_n=(\underline{\underline{R}}_{n}^{i})^{1/2}\underline{I}_n^i$ is a current-like vector that describes the input signal of the cavity $n$. The power entering the $n^{th}$ cavity is given by,
\begin{equation} \label{eq:a7}
    P_n=\frac{1}{2}Re \left[ {\underline{I}_n^i}^{\dagger} \underline{\underline{Z}}_n^{ii} \, \underline{I}_n^i \right] = \frac{1}{2} \underline{\psi}_n^{\dagger} \underline{\psi}_n
\end{equation}
We first assume the transition matrix $\underline{\underline{\tau}}$ to be diagonal. The input power at the $(n+1)^{st}$ cavity is written as: 
\begin{equation} \label{eq:pnj1}
\begin{aligned}
    P_{n+1} &= \frac{1}{2} \underline{\psi}_{n+1}^{\dagger} \underline{\psi}_{n+1} \\
    &=\frac{1}{2} \left( \underline{\underline{\tau}}_{n\rightarrow n+1}\,\underline{\underline{\xi}}_n\,{\underline{\psi}_n} \right)^{\dagger} \left( \underline{\underline{\tau}}_{n\rightarrow n+1}\,\underline{\underline{\xi}}_n\,{\underline{\psi}_n} \right) \\
    &= \frac{1}{2} \sum_{i,j,k} \left(\psi_{n,i}^* \,\xi_{ji}^* \left|{{\tau}}_{j} \right|^2\,{{\xi_{jk}}}\,{{\psi_{n,k}}} \right)
\end{aligned}
\end{equation}
The more general cases, where the constraint of $\underline{\underline{\tau}}$ being diagonal is lifted, will be discussed later in this section.
With $\left< P_n \right> = \frac{1}{2} \left< \underline{\psi}_n^{\dagger} \underline{\psi}_n\right>$ being the average input power at the $n^{th}$ cavity, and the fact that $\left< \xi_{ji}^* {{\xi_{j'i'}}} \right> = \left< \xi_{ji}^* {{\xi_{ji}}} \right>\delta_{ii'}\delta_{jj'}$, we can further simplify Eq. (\ref{eq:pnj1}) as:
\begin{equation} \label{eq:pcascade}
    \left< P_{n+1} \right> = \left< P_{n} \right> \cdot \left< \left|{{\xi}} \right|^2 \right> \cdot \sum_j \left|{{\tau}}_{j} \right|^2 = \left< P_{n} \right> \left< \left|{{\xi}} \right|^2 \right> \sum_j T_j
\end{equation}
where $T_j = \left|{{\tau}}_{j} \right|^2$ and the sum on j runs from all transmitting channels. Based on the relationship in Eq.\ (\ref{eq:pcascade}), we next analyze the fluctuation level of the power delivered to the load ($P_L$) by studying the ratio $\kappa=\left< P_L^2 \right>/\left<P_L\right>^2$. To simplify the following expression, $X_j = \psi_{n,j}^*{\psi_{n,j}}$ and $Y_j = \psi_{n+1,j}^*{\psi_{n+1,j}}$ are employed in the following calculations. 
Equation (\ref{eq:pnj1}) is now rewritten as $Y_j = T_j \sum_{i,i'} \psi_{n,i}^*{\psi_{n,i'}}\, \xi_{ji}^* {{\xi_{ji'}}}$, and further we have $\left< Y_j \right> = T_j \left< \left|{{\xi}} \right|^2 \right> \sum_i \left<X_i\right>$. 
Utilizing $\left< \left|{{\xi}} \right|^4 \right>=2\left< \left|{{\xi}}\right|^2 \right>^2$ and $\left< \xi_{ji}^*{{\xi_{ji'}}}\,
\xi_{ji}^*{{\xi_{ji'}}}\right>=\left< \xi_{ji}^*\xi_{ji}^*\right>\left< {{\xi_{ji'}}}\,{{\xi_{ji'}}}\right>=0$ when $i \neq i'$, we have:
\begin{equation} \label{eq:a8}
    \sum_{jj'}\left<Y_jY_{j'}\right> = \left< \left|{{\xi}}\right|^2 \right>^2 \sum_{i,i''} \left< X_i X_{i''}\right>\left[ \sum_{jj'} T_j T_{j'} + \sum_j T_j^2\right]
\end{equation}
We then introduce a mean power transmission coefficient for the $n^{th}$ cavity $\Bar{T}_n$ and an effective total number of channels $M_n$, defined as:
\begin{equation}
    \Bar{T}_n = \left< \left|{{\xi}}\right|^2 \right> \sum_j T_j, \quad M_n^{-1} = \sum_j T_j^2/\left( \sum_j T_j\right)^2,
\end{equation}
respectively. Thus we have the power transmitted into the $(n+1)^{st}$ cavity (Eq.\ (\ref{eq:pcascade})) written as:
\begin{equation}
    \left<P_{n+1}\right> = \Bar{T}_n \left<P_{n}\right>,
\end{equation}
and
\begin{equation} \label{eq:a11}
    \left<P_{n+1}^2\right> = \Bar{T}_n^2 \left( 1+ M_n^{-1}\right) \left<P_{n}^2\right>.
\end{equation}
The fluctuation level of the power after N cavities is:
\begin{equation}\label{eq:kapa}
    \kappa = \frac{\left< P_L^2 \right>}{\left<P_L\right>^2} = \frac{\left< P_1^2 \right>}{\left<P_1\right>^2} \prod_{n=1}^{N}\left( 1+ M_n^{-1}\right)=\prod_{n=1}^{N}\left( 1+ M_n^{-1}\right) .
\end{equation}

Here we utilized the fact that the power injected into the first cavity $(P_1)$ is a fixed scalar in the case of a steady input.
Since the last cavity is connected to the load with a single-mode port in experiments, we have $\left( 1+ M_N^{-1}\right)= 2$ at the last cavity. 
The overall fluctuating level of load power is 
\begin{equation}\label{eq:a12}
\kappa = \frac{\left< P_L^2 \right>}{\left<P_L\right>^2} = 2 \times \left[ \prod_{n=1}^{N-1}\left( 1+ M_n^{-1}\right) \right].
\end{equation}
For a system with a large number of connecting channels ($M_n\gg 1$) between the neighbouring cavities, such as the circular aperture connection which allows $\sim100$ propagating modes, the factor $(1+M_n^{-1}) \rightarrow 1$. In this limit, the mean field methods are sufficient to describe the power flow for cavities with strong coupling.  On the contrary, the hybrid model is not expected to work for system with few coupling channels inside a cavity cascade. 
The study explains the experimental observations that hybrid models are successfully applied to systems with multi-mode apertures connections (Fig.\ \ref{fig:induceV_a}), while fail to generate predictions for systems with ``bottlenecks'' (Fig. \ref{fig:breakdown} (b)).

We also note that the proposed hybrid model assumes each input current throughout the chain is Gaussian distributed while the RCM predicts non-Gaussian statistics \cite{Gradoni2012}.
In the presence of a ``bottleneck'' connection, a deviation from Gaussian statistics of the input current distribution will be present caused by the fluctuating nature of the narrow aperture coupling.

We next expand the analysis to the more general cases where the transition matrix $\underline{\underline{\tau}}$ is no longer a diagonal matrix. With Eqs.\ (\ref{eq:pnj1}) and (\ref{eq:a8}), the expression for $Y_j$ becomes:
\begin{equation}
    Y_j=\sum_{j'j''}{{\tau}}_{jj'}^*{{\tau}}_{jj''}\sum_{i',i''} \psi_{n,i'}^{*} \psi_{n,i''}\,
    \xi_{j'i'}^* {{\xi_{j''i''}}}.
\end{equation}
And correspondingly the average of $Y_j$ is $\left< Y_j \right> = \sum_{j'} {{\tau}}_{jj'}^*{{\tau}}_{jj'} \left< \left|{{\xi}} \right|^2 \right> \sum_i \left<X_i\right>$. With the updated $\Bar{T}_n=\left< \left|{{\xi}}\right|^2 \right> \sum_{jj'} {{\tau}}_{jj'}^*{{\tau}}_{jj'}$, we sum on $j$ and have
\begin{equation}
    \sum_j\left< Y_j \right>=\left< \left|{{\xi}} \right|^2 \right> \sum_{jj'} {{\tau}}_{jj'}^*{{\tau}}_{jj''} \sum_i \left<X_i\right> = \Bar{T}_n \sum_i \left<X_i\right>,
\end{equation}
and the average of the product of $Y$'s is
\begin{equation} \label{eq:end}
\begin{aligned}
      \left< Y_k Y_{k'} \right> = &\sum_{j,j',j'',j'''}
      \tau_{kj}^*\tau_{kj'}\tau_{k'j''}^*\tau_{k'j'''} \, \cdot \\
      &\sum_{i,i',i'',i'''}\left< \psi_{n,i}^*\psi_{n,i'}\psi_{n,i''}^*\psi_{n,i'''} \right>\left< \xi_{ji}^*\xi_{j'i'}\xi_{j''i''}^*\xi_{j'''i'''} \right>.  
\end{aligned}
\end{equation}
With careful calculation, the double summation of Eq. (\ref{eq:end}) over $k$ and $k'$ gives:
\begin{equation}
\begin{aligned}
    \sum_{k,k'}\left< Y_k Y_{k'} \right> / \left< \left|{{\xi}}\right|^2 \right>^2 = &\sum_{ii'} \left<X_i X_{i'}\right> \left[ \sum_{jk} {{\tau}}_{kj}^*{{\tau}}_{kj} \right]^2 + \\
    &\sum_{ii'} \left<X_i X_{i'}\right> \sum_{jj'} \left[ \sum_{k} {{\tau}}_{kj}^*{{\tau}}_{kj'} \right]^2.
\end{aligned}
\end{equation}
With the updated $\Bar{T}_n$, we reach a similar expression of the power flow as in Eq. (\ref{eq:a11}) with the new $M_n^{-1}$ written as
\begin{equation}
    M_n^{-1}=\sum_{jj'} \left[ \sum_{k} {{\tau}}_{kj}^*{{\tau}}_{kj'} \right]^2 / \left[\sum_{jj'} {{\tau}}_{jj'}^*{{\tau}}_{jj'} \right]^2.
\end{equation}
Thus the conclusions of the multi-channel analysis can be applied to the more general cases with the refined $\underline{\underline{\tau}}$, $\Bar{T}_n$ and $M_n^{-1}$.

\section{Appendix D: Applying the Hybrid Model to Generic Cavity Systems}

\begin{figure}
\includegraphics[width=0.5\textwidth]{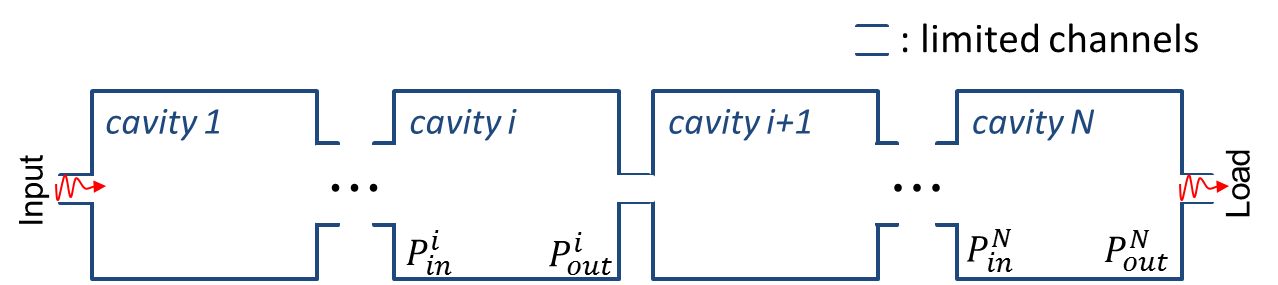}
\caption{\label{fig:choke} Schematic of a cavity cascade array with many-channel connecting apertures between all cavities except that between cavity $i$ and cavity $i+1$. Note the definition of input and output powers at each cavity aperture. }
\end{figure}

Here we propose a scheme of applying the hybrid model to more general cases, where there exists several limited-channel connections between the intermediate cavities. Due to the large power flow fluctuations induced by these ``bottlenecks'', we apply RCM to better characterize these connections. 
A general case is shown in Fig.\ \ref{fig:choke}. We study the treatment of limited channel connections between the $i^{th}$ and $(i+1)^{st}$ cavities in a N-cavity cascade chain. The treatment involves four steps:
\begin{enumerate}

\item Use the PWB N-cavity model to calculate the input, output power of the $i^{th}$ cavity $P_{in}^i |_{PWB}$, $P_{out}^i |_{PWB}$, and the input power at the last ($N^{th}$) cavity $P_{in}^N |_{PWB}$.

\item For the case of a bottleneck between cavity $i$ and cavity $i+1$ (Fig.\ \ref{fig:choke}), use the calculated $P_{in}^i |_{PWB}$ and RCM treatment to calculate an ensemble of the output power of the $i^{th}$ cavity $P_{out}^i |_{RCM}$.

\item Update the input power at the last cavity by 
\[ P_{in}^N |_{RCM} = P_{out}^i |_{RCM} \cdot \frac{P_{in}^N}{P_{out}^i}|_{PWB}.\]

 Note that we therefore obtain an ensemble of the input power to the last cavity $P_{in}^N |_{RCM}$.

\item Use RCM to calculate an ensemble of the output power of the $N^{th}$ cavity $P_{out}^N |_{RCM}$ with the ensemble of $P_{in}^N |_{RCM}$ obtained from (3). 
\end{enumerate}

For cases where there exist multiple ``bottlenecks'' inside the chain (beyond cavity $i$), one may repeat the procedure (2)-(4) by actively changing the cavity index $N$ to the next cavity connected through a bottleneck.

\section{Appendix E: Hybrid Model Simplification}

\begin{figure}
\includegraphics[width=0.5\textwidth]{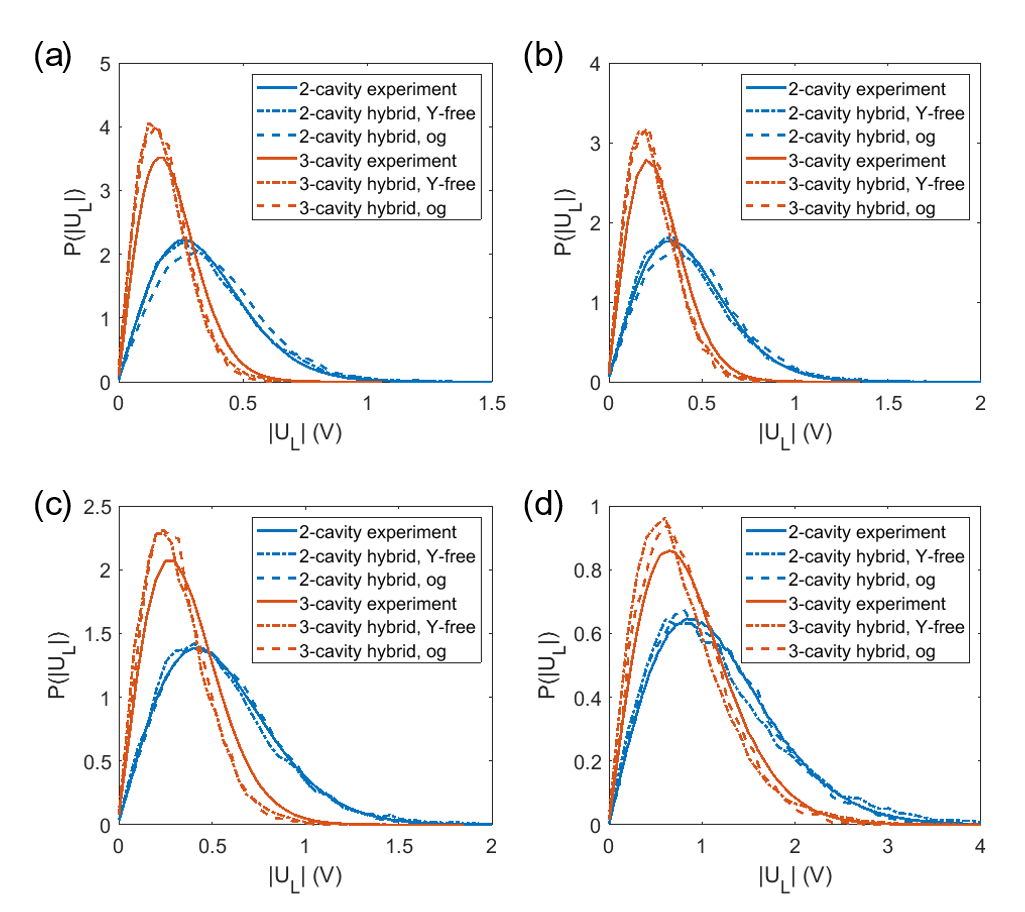}
\caption{\label{fig:simplify} The PDFs of load induced voltage $|U_L|$ of 2- and 3-cavity experiments (solid), hybrid model (dashed), and the simplified hybrid model (dash-dotted). The single cavity loss parameter is varied from 9.7, 7.5, 5.7 and 1.7 from (a-d), respectively. }
\end{figure}

Here we present a simplified treatment of the PWB-RCM hybrid model where the full-wave simulated aperture radiation admittance matrix is no longer required. We assume that all inter-cavity apertures are large on the scale of a wavelength, ie.\ the original PWB-RCM treatment is applicable. 
We also will assume that all cavities are in the high-loss ($\alpha>1$) limit.
This $Y_{aper}$-free treatment will further decrease the computational cost of the hybrid model.
We utilize Eqs. (\ref{eq:a6}-\ref{eq:pnj1}) from Appendix C to introduce the new treatment.

Consider an N-cavity cascade chain, one may treat the load connected to the output port of the $N^{th}$ cavity as the $(N+1)^{st}$ effective cavity. 
The induced power is equal to the input power of the $(N+1)^{st}$ cavity:
\begin{equation} \label{eq:ddd}
\begin{aligned}
    P_{load} = P_{N+1} &= \frac{1}{2} {\psi_{N+1}}^{\dagger} {\psi_{N+1}} \\
    &=\frac{1}{2} \left( {{\tau}}_{N\rightarrow N+1}\,{\underline{\xi}_N}\,{\underline{\psi}_n} \right)^{\dagger} \left( {{\tau}}_{N\rightarrow N+1}\,{\underline{\xi}_N}\,{\underline{\psi}_n} \right) 
\end{aligned}
\end{equation}
where $\tau_{N\rightarrow N+1} = ({{R_{N+1}^{i}}})^{1/2}({{Z_{N+1}^{ii}}}+{{Z_N^{oo}}})^{-1} ({{R_{N}^{o}}})^{1/2}$ is a scalar transition factor due to the fact that a single-mode port acts as a single-mode `aperture' connecting the $N^{th}$ and $(N+1)^{st}$ cavity (the load).
As introduced in the main text, the radiation impedance of the port, which is used to calculate $\tau_{N\rightarrow N+1}$, is obtained from the scattering matrix measurements of the waveguide ports.

Equation (\ref{eq:ddd}) is further generalized as:
\begin{equation} \label{eq:ddd2}
    P_{load} = \frac{1}{2} {\psi_{N+1}}^{\dagger} {\psi_{N+1}}=\frac{1}{2} \left| \tau_{N\rightarrow N+1} \right|^2 \underline{\psi}_n^\dagger \,\underline{\xi}_N^\dagger \,\underline{\xi}_N \,\underline{\psi}_n.
\end{equation}
The ensemble average of $P_{load}$ is
\begin{equation} \label{eq:avg}
\begin{aligned}
    \left< P_{load} \right> &= \frac{1}{2} \left| \tau_{N\rightarrow N+1} \right|^2 \left< \underline{\psi}_n^\dagger \,\underline{\xi}_N^\dagger \,\underline{\xi}_N \,\underline{\psi}_n \right> \\
    &= \frac{1}{2} \left| \tau_{N\rightarrow N+1} \right|^2 \underline{\psi}_n^\dagger \underline{\psi}_n \left< \underline{\xi}_N^\dagger \underline{\xi}_N \right>.
\end{aligned}
\end{equation}

In the above equation, the quantity $\underline{\psi}_n$ is moved out of the ensemble averaging because only $\underline{\xi}_N$ is varying under ensemble average while $\underline{\psi}_n$ and $\tau_{N\rightarrow N+1}$ are not. 
As defined in Appendix C, we write $\underline{\psi}_n={\underline{\underline{R}}_N^{ii}}^{1/2} \cdot \underline{I}_n^i$ under the high-loss assumption. 
Here $\underline{\underline{R}}_N^{ii}$ is set as the radiation impedance of the aperture which does not change value under ensemble averaging.
The input current vector $\underline{I}_n^i$ is also considered to be not changing since the cavity off-diagonal impedance element $\underline{\underline{Z}}_n^{io}$ (Eq.\ (\ref{eq:cav})) is small at high-loss limit, and the input power of the $N^{th}$ cavity is fixed from the prior PWB calculation.
Since the input power of the $N^{th}$ cavity $P_N=\frac{1}{2} \underline{\psi}_{N}^{\dagger} \underline{\psi}_{N}$ is given by the PWB calculation, we may further simplify Eq.\ (\ref{eq:avg}) as $\left< P_{load} \right> = \frac{1}{2} \left| \tau_{N\rightarrow N+1} \right|^2 \underline{\psi}_n^\dagger \underline{\psi}_n \left< \underline{\xi}_N^\dagger \underline{\xi}_N \right> = P_N \left| \tau_{N\rightarrow N+1} \right|^2 \cdot \left< \underline{\xi}_N^\dagger \underline{\xi}_N \right>$.
The vector element of $\underline{\xi}_N$, $\xi_i$, is generated using the same method as in Section III.
The vector product $\underline{\xi}_N^\dagger \underline{\xi}_N$ is the summation of a large number of $\xi_i^* \xi_i$s.
Thus $\underline{\xi}_N^\dagger \underline{\xi}_N$ and $\xi_i^* \xi_i$ have the same expected value according to the law of large numbers.
We finally write $\left< P_{load} \right> = P_N \left| \tau_{N\rightarrow N+1} \right|^2 \left< \xi_i^* \xi_i \right>$ and Eq. (\ref{eq:ddd2}) is reduced to scalar components $P_{load} = P_N \left| \tau_{N\rightarrow N+1} \right|^2 \xi_i^* \xi_i$.

The comparison of the original hybrid model and the $Y_{aper}$-free hybrid model is presented in Fig. \ref{fig:simplify}.
Good agreement between the simplified hybrid model predictions and the experimental cases for the PDF of load induced voltages are found.
The original hybrid model slightly outperforms its simplified version for the 3-cavity cases (red curves in Fig. \ref{fig:simplify} (a-d)), and it is expected that the accuracy of predictions will decrease for longer chains.
Such an effect may be caused by the absence of an aperture admittance matrix in the simplified hybrid model.
The $Y_{aper}$-free treatment would broaden the applicable range of the PWB-RCM hybrid model to situations where the exact shape of the aperture is unknown and the numerical simulation of the aperture is unavailable.
However, this shortcut shares the same high-loss ($\alpha>1$) assumption since it is built on the multi-channel analysis (Appendix C).
We also note that for situations discussed in Appendix D, the admittance matrix of the narrow aperture (bottleneck) is required to compute the transition matrix $\underline{\underline{\tau}}$.

\bibliographystyle{apsrev4-1} 

\begin{thebibliography}{65}%
\makeatletter
\providecommand \@ifxundefined [1]{%
 \@ifx{#1\undefined}
}%
\providecommand \@ifnum [1]{%
 \ifnum #1\expandafter \@firstoftwo
 \else \expandafter \@secondoftwo
 \fi
}%
\providecommand \@ifx [1]{%
 \ifx #1\expandafter \@firstoftwo
 \else \expandafter \@secondoftwo
 \fi
}%
\providecommand \natexlab [1]{#1}%
\providecommand \enquote  [1]{``#1''}%
\providecommand \bibnamefont  [1]{#1}%
\providecommand \bibfnamefont [1]{#1}%
\providecommand \citenamefont [1]{#1}%
\providecommand \href@noop [0]{\@secondoftwo}%
\providecommand \href [0]{\begingroup \@sanitize@url \@href}%
\providecommand \@href[1]{\@@startlink{#1}\@@href}%
\providecommand \@@href[1]{\endgroup#1\@@endlink}%
\providecommand \@sanitize@url [0]{\catcode `\\12\catcode `\$12\catcode
  `\&12\catcode `\#12\catcode `\^12\catcode `\_12\catcode `\%12\relax}%
\providecommand \@@startlink[1]{}%
\providecommand \@@endlink[0]{}%
\providecommand \url  [0]{\begingroup\@sanitize@url \@url }%
\providecommand \@url [1]{\endgroup\@href {#1}{\urlprefix }}%
\providecommand \urlprefix  [0]{URL }%
\providecommand \Eprint [0]{\href }%
\providecommand \doibase [0]{http://dx.doi.org/}%
\providecommand \selectlanguage [0]{\@gobble}%
\providecommand \bibinfo  [0]{\@secondoftwo}%
\providecommand \bibfield  [0]{\@secondoftwo}%
\providecommand \translation [1]{[#1]}%
\providecommand \BibitemOpen [0]{}%
\providecommand \bibitemStop [0]{}%
\providecommand \bibitemNoStop [0]{.\EOS\space}%
\providecommand \EOS [0]{\spacefactor3000\relax}%
\providecommand \BibitemShut  [1]{\csname bibitem#1\endcsname}%
\let\auto@bib@innerbib\@empty
\bibitem [{\citenamefont {Hemmady}\ \emph {et~al.}(2005)\citenamefont
  {Hemmady}, \citenamefont {Zheng}, \citenamefont {Ott}, \citenamefont
  {Antonsen},\ and\ \citenamefont {Anlage}}]{Hemmady2005}%
  \BibitemOpen
  \bibfield  {author} {\bibinfo {author} {\bibfnamefont {S.}~\bibnamefont
  {Hemmady}}, \bibinfo {author} {\bibfnamefont {X.}~\bibnamefont {Zheng}},
  \bibinfo {author} {\bibfnamefont {E.}~\bibnamefont {Ott}}, \bibinfo {author}
  {\bibfnamefont {T.~M.}\ \bibnamefont {Antonsen}}, \ and\ \bibinfo {author}
  {\bibfnamefont {S.~M.}\ \bibnamefont {Anlage}},\ }\href {\doibase
  10.1103/PhysRevLett.94.014102} {\bibfield  {journal} {\bibinfo  {journal}
  {Physical Review Letters}\ }\textbf {\bibinfo {volume} {94}},\ \bibinfo
  {pages} {014102} (\bibinfo {year} {2005})}\BibitemShut {NoStop}%
\bibitem [{\citenamefont {Zheng}\ \emph
  {et~al.}(2006{\natexlab{a}})\citenamefont {Zheng}, \citenamefont {Antonsen},\
  and\ \citenamefont {Ott}}]{Zheng2006}%
  \BibitemOpen
  \bibfield  {author} {\bibinfo {author} {\bibfnamefont {X.}~\bibnamefont
  {Zheng}}, \bibinfo {author} {\bibfnamefont {T.~M.}\ \bibnamefont {Antonsen}},
  \ and\ \bibinfo {author} {\bibfnamefont {E.}~\bibnamefont {Ott}},\ }\href
  {\doibase 10.1080/02726340500214902} {\bibfield  {journal} {\bibinfo
  {journal} {Electromagnetics}\ }\textbf {\bibinfo {volume} {26}},\ \bibinfo
  {pages} {37} (\bibinfo {year} {2006}{\natexlab{a}})}\BibitemShut {NoStop}%
\bibitem [{\citenamefont {Zheng}\ \emph
  {et~al.}(2006{\natexlab{b}})\citenamefont {Zheng}, \citenamefont {Antonsen},\
  and\ \citenamefont {Ott}}]{Zheng2006a}%
  \BibitemOpen
  \bibfield  {author} {\bibinfo {author} {\bibfnamefont {X.}~\bibnamefont
  {Zheng}}, \bibinfo {author} {\bibfnamefont {T.~M.}\ \bibnamefont {Antonsen}},
  \ and\ \bibinfo {author} {\bibfnamefont {E.}~\bibnamefont {Ott}},\ }\href
  {\doibase 10.1080/02726340500214894} {\bibfield  {journal} {\bibinfo
  {journal} {Electromagnetics}\ }\textbf {\bibinfo {volume} {26}},\ \bibinfo
  {pages} {3} (\bibinfo {year} {2006}{\natexlab{b}})}\BibitemShut {NoStop}%
\bibitem [{\citenamefont {Hemmady}\ \emph {et~al.}(2012)\citenamefont
  {Hemmady}, \citenamefont {Antonsen}, \citenamefont {Ott},\ and\ \citenamefont
  {Anlage}}]{Hemmady2012}%
  \BibitemOpen
  \bibfield  {author} {\bibinfo {author} {\bibfnamefont {S.}~\bibnamefont
  {Hemmady}}, \bibinfo {author} {\bibfnamefont {T.~M.}\ \bibnamefont
  {Antonsen}}, \bibinfo {author} {\bibfnamefont {E.}~\bibnamefont {Ott}}, \
  and\ \bibinfo {author} {\bibfnamefont {S.~M.}\ \bibnamefont {Anlage}},\
  }\href {\doibase 10.1109/TEMC.2011.2177270} {\bibfield  {journal} {\bibinfo
  {journal} {IEEE Transactions on Electromagnetic Compatibility}\ }\textbf
  {\bibinfo {volume} {54}},\ \bibinfo {pages} {758} (\bibinfo {year}
  {2012})}\BibitemShut {NoStop}%
\bibitem [{\citenamefont {Li}\ \emph {et~al.}(2015)\citenamefont {Li},
  \citenamefont {Meng}, \citenamefont {Liu}, \citenamefont {Schamiloglu},\ and\
  \citenamefont {Hemmady}}]{Li2015}%
  \BibitemOpen
  \bibfield  {author} {\bibinfo {author} {\bibfnamefont {X.}~\bibnamefont
  {Li}}, \bibinfo {author} {\bibfnamefont {C.}~\bibnamefont {Meng}}, \bibinfo
  {author} {\bibfnamefont {Y.}~\bibnamefont {Liu}}, \bibinfo {author}
  {\bibfnamefont {E.}~\bibnamefont {Schamiloglu}}, \ and\ \bibinfo {author}
  {\bibfnamefont {S.~D.}\ \bibnamefont {Hemmady}},\ }\href {\doibase
  10.1109/TEMC.2014.2384482} {\bibfield  {journal} {\bibinfo  {journal} {IEEE
  Transactions on Electromagnetic Compatibility}\ }\textbf {\bibinfo {volume}
  {57}},\ \bibinfo {pages} {448} (\bibinfo {year} {2015})}\BibitemShut
  {NoStop}%
\bibitem [{\citenamefont {Fedeli}\ \emph {et~al.}(2009)\citenamefont {Fedeli},
  \citenamefont {Gradoni}, \citenamefont {Primiani},\ and\ \citenamefont
  {Moglie}}]{Fedeli2009}%
  \BibitemOpen
  \bibfield  {author} {\bibinfo {author} {\bibfnamefont {D.}~\bibnamefont
  {Fedeli}}, \bibinfo {author} {\bibfnamefont {G.}~\bibnamefont {Gradoni}},
  \bibinfo {author} {\bibfnamefont {V.~M.}\ \bibnamefont {Primiani}}, \ and\
  \bibinfo {author} {\bibfnamefont {F.}~\bibnamefont {Moglie}},\ }\href
  {\doibase 10.1109/TEMC.2009.2013456} {\bibfield  {journal} {\bibinfo
  {journal} {IEEE Transactions on Electromagnetic Compatibility}\ }\textbf
  {\bibinfo {volume} {51}},\ \bibinfo {pages} {170} (\bibinfo {year}
  {2009})}\BibitemShut {NoStop}%
\bibitem [{\citenamefont {Dietz}\ \emph {et~al.}(2006)\citenamefont {Dietz},
  \citenamefont {Guhr}, \citenamefont {Harney},\ and\ \citenamefont
  {Richter}}]{Dietz2006}%
  \BibitemOpen
  \bibfield  {author} {\bibinfo {author} {\bibfnamefont {B.}~\bibnamefont
  {Dietz}}, \bibinfo {author} {\bibfnamefont {T.}~\bibnamefont {Guhr}},
  \bibinfo {author} {\bibfnamefont {H.~L.}\ \bibnamefont {Harney}}, \ and\
  \bibinfo {author} {\bibfnamefont {A.}~\bibnamefont {Richter}},\ }\href
  {\doibase 10.1103/PhysRevLett.96.254101} {\bibfield  {journal} {\bibinfo
  {journal} {Physical Review Letters}\ }\textbf {\bibinfo {volume} {96}},\
  \bibinfo {pages} {254101} (\bibinfo {year} {2006})}\BibitemShut {NoStop}%
\bibitem [{\citenamefont {Chalker}\ and\ \citenamefont
  {Coddington}(1988)}]{Chalker1988}%
  \BibitemOpen
  \bibfield  {author} {\bibinfo {author} {\bibfnamefont {J.~T.}\ \bibnamefont
  {Chalker}}\ and\ \bibinfo {author} {\bibfnamefont {P.~D.}\ \bibnamefont
  {Coddington}},\ }\href {\doibase 10.1088/0022-3719/21/14/008} {\bibfield
  {journal} {\bibinfo  {journal} {Journal of Physics C: Solid State Physics}\
  }\textbf {\bibinfo {volume} {21}},\ \bibinfo {pages} {2665} (\bibinfo {year}
  {1988})}\BibitemShut {NoStop}%
\bibitem [{\citenamefont {Sau}\ and\ \citenamefont {Sarma}(2012)}]{Sau2012}%
  \BibitemOpen
  \bibfield  {author} {\bibinfo {author} {\bibfnamefont {J.~D.}\ \bibnamefont
  {Sau}}\ and\ \bibinfo {author} {\bibfnamefont {S.~D.}\ \bibnamefont
  {Sarma}},\ }\href {\doibase 10.1038/ncomms1966} {\bibfield  {journal}
  {\bibinfo  {journal} {Nature Communications}\ }\textbf {\bibinfo {volume}
  {3}},\ \bibinfo {pages} {964} (\bibinfo {year} {2012})}\BibitemShut {NoStop}%
\bibitem [{\citenamefont {Beenakker}(2015)}]{Beenakker2015}%
  \BibitemOpen
  \bibfield  {author} {\bibinfo {author} {\bibfnamefont {C.~W.~J.}\
  \bibnamefont {Beenakker}},\ }\href {\doibase 10.1103/RevModPhys.87.1037}
  {\bibfield  {journal} {\bibinfo  {journal} {Reviews of Modern Physics}\
  }\textbf {\bibinfo {volume} {87}},\ \bibinfo {pages} {1037} (\bibinfo {year}
  {2015})}\BibitemShut {NoStop}%
\bibitem [{\citenamefont {Zhang}\ and\ \citenamefont {Nori}(2016)}]{Zhang}%
  \BibitemOpen
  \bibfield  {author} {\bibinfo {author} {\bibfnamefont {P.}~\bibnamefont
  {Zhang}}\ and\ \bibinfo {author} {\bibfnamefont {F.}~\bibnamefont {Nori}},\
  }\href {\doibase 10.1088/1367-2630/18/4/043033} {\bibfield  {journal}
  {\bibinfo  {journal} {New Journal of Physics}\ }\textbf {\bibinfo {volume}
  {18}},\ \bibinfo {pages} {043033} (\bibinfo {year} {2016})}\BibitemShut
  {NoStop}%
\bibitem [{\citenamefont {Dupr{\'{e}}}\ \emph {et~al.}(2015)\citenamefont
  {Dupr{\'{e}}}, \citenamefont {{Del Hougne}}, \citenamefont {Fink},
  \citenamefont {Lemoult},\ and\ \citenamefont {Lerosey}}]{Dupre2015}%
  \BibitemOpen
  \bibfield  {author} {\bibinfo {author} {\bibfnamefont {M.}~\bibnamefont
  {Dupr{\'{e}}}}, \bibinfo {author} {\bibfnamefont {P.}~\bibnamefont {{Del
  Hougne}}}, \bibinfo {author} {\bibfnamefont {M.}~\bibnamefont {Fink}},
  \bibinfo {author} {\bibfnamefont {F.}~\bibnamefont {Lemoult}}, \ and\
  \bibinfo {author} {\bibfnamefont {G.}~\bibnamefont {Lerosey}},\ }\href
  {\doibase 10.1103/PhysRevLett.115.017701} {\bibfield  {journal} {\bibinfo
  {journal} {Physical Review Letters}\ }\textbf {\bibinfo {volume} {115}},\
  \bibinfo {pages} {017701} (\bibinfo {year} {2015})}\BibitemShut {NoStop}%
\bibitem [{\citenamefont {Kaina}\ \emph {et~al.}(2015)\citenamefont {Kaina},
  \citenamefont {Dupr{\'{e}}}, \citenamefont {Lerosey},\ and\ \citenamefont
  {Fink}}]{Kaina2015}%
  \BibitemOpen
  \bibfield  {author} {\bibinfo {author} {\bibfnamefont {N.}~\bibnamefont
  {Kaina}}, \bibinfo {author} {\bibfnamefont {M.}~\bibnamefont {Dupr{\'{e}}}},
  \bibinfo {author} {\bibfnamefont {G.}~\bibnamefont {Lerosey}}, \ and\
  \bibinfo {author} {\bibfnamefont {M.}~\bibnamefont {Fink}},\ }\href {\doibase
  10.1038/srep06693} {\bibfield  {journal} {\bibinfo  {journal} {Scientific
  Reports}\ }\textbf {\bibinfo {volume} {4}},\ \bibinfo {pages} {6693}
  (\bibinfo {year} {2015})}\BibitemShut {NoStop}%
\bibitem [{\citenamefont {del Hougne}\ \emph {et~al.}(2018)\citenamefont {del
  Hougne}, \citenamefont {{F. Imani}}, \citenamefont {Sleasman}, \citenamefont
  {Gollub}, \citenamefont {Fink}, \citenamefont {Lerosey},\ and\ \citenamefont
  {Smith}}]{DelHougne2018}%
  \BibitemOpen
  \bibfield  {author} {\bibinfo {author} {\bibfnamefont {P.}~\bibnamefont {del
  Hougne}}, \bibinfo {author} {\bibfnamefont {M.}~\bibnamefont {{F. Imani}}},
  \bibinfo {author} {\bibfnamefont {T.}~\bibnamefont {Sleasman}}, \bibinfo
  {author} {\bibfnamefont {J.~N.}\ \bibnamefont {Gollub}}, \bibinfo {author}
  {\bibfnamefont {M.}~\bibnamefont {Fink}}, \bibinfo {author} {\bibfnamefont
  {G.}~\bibnamefont {Lerosey}}, \ and\ \bibinfo {author} {\bibfnamefont
  {D.~R.}\ \bibnamefont {Smith}},\ }\href {\doibase 10.1038/s41598-018-24681-9}
  {\bibfield  {journal} {\bibinfo  {journal} {Scientific Reports}\ }\textbf
  {\bibinfo {volume} {8}},\ \bibinfo {pages} {6536} (\bibinfo {year}
  {2018})}\BibitemShut {NoStop}%
\bibitem [{\citenamefont {Ellegaard}\ \emph {et~al.}(1995)\citenamefont
  {Ellegaard}, \citenamefont {Guhr}, \citenamefont {Lindemann}, \citenamefont
  {Lorensen}, \citenamefont {Nyg{\aa}rd},\ and\ \citenamefont
  {Oxborrow}}]{Ellegaard1995}%
  \BibitemOpen
  \bibfield  {author} {\bibinfo {author} {\bibfnamefont {C.}~\bibnamefont
  {Ellegaard}}, \bibinfo {author} {\bibfnamefont {T.}~\bibnamefont {Guhr}},
  \bibinfo {author} {\bibfnamefont {K.}~\bibnamefont {Lindemann}}, \bibinfo
  {author} {\bibfnamefont {H.~Q.}\ \bibnamefont {Lorensen}}, \bibinfo {author}
  {\bibfnamefont {J.}~\bibnamefont {Nyg{\aa}rd}}, \ and\ \bibinfo {author}
  {\bibfnamefont {M.}~\bibnamefont {Oxborrow}},\ }\href {\doibase
  10.1103/PhysRevLett.75.1546} {\enquote {\bibinfo {title} {{Spectral
  Statistics of Acoustic Resonances in Aluminum Blocks}},}\ } (\bibinfo {year}
  {1995})\BibitemShut {NoStop}%
\bibitem [{\citenamefont {Tanner}\ and\ \citenamefont
  {S{\o}ndergaard}(2007)}]{TS07}%
  \BibitemOpen
  \bibfield  {author} {\bibinfo {author} {\bibfnamefont {G.}~\bibnamefont
  {Tanner}}\ and\ \bibinfo {author} {\bibfnamefont {N.}~\bibnamefont
  {S{\o}ndergaard}},\ }\href@noop {} {\bibfield  {journal} {\bibinfo  {journal}
  {J. Phys. A}\ }\textbf {\bibinfo {volume} {40}},\ \bibinfo {pages} {R443}
  (\bibinfo {year} {2007})}\BibitemShut {NoStop}%
\bibitem [{\citenamefont {Aur{\'{e}}gan}\ and\ \citenamefont
  {Pagneux}(2016)}]{Auregan2016}%
  \BibitemOpen
  \bibfield  {author} {\bibinfo {author} {\bibfnamefont {Y.}~\bibnamefont
  {Aur{\'{e}}gan}}\ and\ \bibinfo {author} {\bibfnamefont {V.}~\bibnamefont
  {Pagneux}},\ }\href {\doibase 10.3813/AAA.919001} {\bibfield  {journal}
  {\bibinfo  {journal} {Acta Acustica united with Acustica}\ }\textbf {\bibinfo
  {volume} {102}},\ \bibinfo {pages} {869} (\bibinfo {year}
  {2016})}\BibitemShut {NoStop}%
\bibitem [{\citenamefont {Doron}\ \emph {et~al.}(1990)\citenamefont {Doron},
  \citenamefont {Smilansky},\ and\ \citenamefont {Frenkel}}]{Doron1990}%
  \BibitemOpen
  \bibfield  {author} {\bibinfo {author} {\bibfnamefont {E.}~\bibnamefont
  {Doron}}, \bibinfo {author} {\bibfnamefont {U.}~\bibnamefont {Smilansky}}, \
  and\ \bibinfo {author} {\bibfnamefont {A.}~\bibnamefont {Frenkel}},\ }\href
  {\doibase 10.1103/PhysRevLett.65.3072} {\bibfield  {journal} {\bibinfo
  {journal} {Physical Review Letters}\ }\textbf {\bibinfo {volume} {65}},\
  \bibinfo {pages} {3072} (\bibinfo {year} {1990})}\BibitemShut {NoStop}%
\bibitem [{\citenamefont {So}\ \emph {et~al.}(1995)\citenamefont {So},
  \citenamefont {Anlage}, \citenamefont {Ott},\ and\ \citenamefont
  {Oerter}}]{So1995}%
  \BibitemOpen
  \bibfield  {author} {\bibinfo {author} {\bibfnamefont {P.}~\bibnamefont
  {So}}, \bibinfo {author} {\bibfnamefont {S.~M.}\ \bibnamefont {Anlage}},
  \bibinfo {author} {\bibfnamefont {E.}~\bibnamefont {Ott}}, \ and\ \bibinfo
  {author} {\bibfnamefont {R.~N.}\ \bibnamefont {Oerter}},\ }\href {\doibase
  10.1103/PhysRevLett.74.2662} {\bibfield  {journal} {\bibinfo  {journal}
  {Physical Review Letters}\ }\textbf {\bibinfo {volume} {74}},\ \bibinfo
  {pages} {2662} (\bibinfo {year} {1995})}\BibitemShut {NoStop}%
\bibitem [{\citenamefont {Kuhl}\ \emph {et~al.}(2005)\citenamefont {Kuhl},
  \citenamefont {Mart{\'{i}}nez-Mares}, \citenamefont
  {M{\'{e}}ndez-S{\'{a}}nchez},\ and\ \citenamefont
  {St{\"{o}}ckmann}}]{Kuhl2005}%
  \BibitemOpen
  \bibfield  {author} {\bibinfo {author} {\bibfnamefont {U.}~\bibnamefont
  {Kuhl}}, \bibinfo {author} {\bibfnamefont {M.}~\bibnamefont
  {Mart{\'{i}}nez-Mares}}, \bibinfo {author} {\bibfnamefont {R.~A.}\
  \bibnamefont {M{\'{e}}ndez-S{\'{a}}nchez}}, \ and\ \bibinfo {author}
  {\bibfnamefont {H.-J.}\ \bibnamefont {St{\"{o}}ckmann}},\ }\href {\doibase
  10.1103/PhysRevLett.94.144101} {\bibfield  {journal} {\bibinfo  {journal}
  {Physical Review Letters}\ }\textbf {\bibinfo {volume} {94}},\ \bibinfo
  {pages} {144101} (\bibinfo {year} {2005})}\BibitemShut {NoStop}%
\bibitem [{\citenamefont {Gradoni}\ \emph {et~al.}(2012)\citenamefont
  {Gradoni}, \citenamefont {Antonsen},\ and\ \citenamefont
  {Ott}}]{Gradoni2012}%
  \BibitemOpen
  \bibfield  {author} {\bibinfo {author} {\bibfnamefont {G.}~\bibnamefont
  {Gradoni}}, \bibinfo {author} {\bibfnamefont {T.~M.}\ \bibnamefont
  {Antonsen}}, \ and\ \bibinfo {author} {\bibfnamefont {E.}~\bibnamefont
  {Ott}},\ }\href {\doibase 10.1103/PhysRevE.86.046204} {\bibfield  {journal}
  {\bibinfo  {journal} {Physical Review E}\ }\textbf {\bibinfo {volume} {86}},\
  \bibinfo {pages} {046204} (\bibinfo {year} {2012})}\BibitemShut {NoStop}%
\bibitem [{\citenamefont {Gagliardi}\ \emph {et~al.}(2015)\citenamefont
  {Gagliardi}, \citenamefont {Micheli}, \citenamefont {Gradoni}, \citenamefont
  {Moglie},\ and\ \citenamefont {Primiani}}]{Gagliardi2015}%
  \BibitemOpen
  \bibfield  {author} {\bibinfo {author} {\bibfnamefont {L.}~\bibnamefont
  {Gagliardi}}, \bibinfo {author} {\bibfnamefont {D.}~\bibnamefont {Micheli}},
  \bibinfo {author} {\bibfnamefont {G.}~\bibnamefont {Gradoni}}, \bibinfo
  {author} {\bibfnamefont {F.}~\bibnamefont {Moglie}}, \ and\ \bibinfo {author}
  {\bibfnamefont {V.~M.}\ \bibnamefont {Primiani}},\ }\href {\doibase
  10.1109/LAWP.2015.2411621} {\bibfield  {journal} {\bibinfo  {journal} {IEEE
  Antennas and Wireless Propagation Letters}\ }\textbf {\bibinfo {volume}
  {14}},\ \bibinfo {pages} {1463} (\bibinfo {year} {2015})}\BibitemShut
  {NoStop}%
\bibitem [{\citenamefont {Xiao}\ \emph {et~al.}(2018)\citenamefont {Xiao},
  \citenamefont {Antonsen}, \citenamefont {Ott}, \citenamefont {Drikas},
  \citenamefont {Gil},\ and\ \citenamefont {Anlage}}]{Xiao2018}%
  \BibitemOpen
  \bibfield  {author} {\bibinfo {author} {\bibfnamefont {B.}~\bibnamefont
  {Xiao}}, \bibinfo {author} {\bibfnamefont {T.~M.}\ \bibnamefont {Antonsen}},
  \bibinfo {author} {\bibfnamefont {E.}~\bibnamefont {Ott}}, \bibinfo {author}
  {\bibfnamefont {Z.~B.}\ \bibnamefont {Drikas}}, \bibinfo {author}
  {\bibfnamefont {J.~G.}\ \bibnamefont {Gil}}, \ and\ \bibinfo {author}
  {\bibfnamefont {S.~M.}\ \bibnamefont {Anlage}},\ }\href {\doibase
  10.1103/PhysRevE.97.062220} {\bibfield  {journal} {\bibinfo  {journal}
  {Physical Review E}\ }\textbf {\bibinfo {volume} {97}},\ \bibinfo {pages}
  {062220} (\bibinfo {year} {2018})}\BibitemShut {NoStop}%
\bibitem [{\citenamefont {Ma}\ \emph {et~al.}(2019)\citenamefont {Ma},
  \citenamefont {Xiao}, \citenamefont {Hong}, \citenamefont {Addissie},
  \citenamefont {Drikas}, \citenamefont {Antonsen}, \citenamefont {Ott},\ and\
  \citenamefont {Anlage}}]{Ma2019}%
  \BibitemOpen
  \bibfield  {author} {\bibinfo {author} {\bibfnamefont {S.}~\bibnamefont
  {Ma}}, \bibinfo {author} {\bibfnamefont {B.}~\bibnamefont {Xiao}}, \bibinfo
  {author} {\bibfnamefont {R.}~\bibnamefont {Hong}}, \bibinfo {author}
  {\bibfnamefont {B.}~\bibnamefont {Addissie}}, \bibinfo {author}
  {\bibfnamefont {Z.}~\bibnamefont {Drikas}}, \bibinfo {author} {\bibfnamefont
  {T.}~\bibnamefont {Antonsen}}, \bibinfo {author} {\bibfnamefont
  {E.}~\bibnamefont {Ott}}, \ and\ \bibinfo {author} {\bibfnamefont
  {S.}~\bibnamefont {Anlage}},\ }\href {\doibase 10.12693/APhysPolA.136.757}
  {\bibfield  {journal} {\bibinfo  {journal} {Acta Physica Polonica A}\
  }\textbf {\bibinfo {volume} {136}},\ \bibinfo {pages} {757} (\bibinfo {year}
  {2019})}\BibitemShut {NoStop}%
\bibitem [{\citenamefont {Tanner}\ \emph {et~al.}(2000)\citenamefont {Tanner},
  \citenamefont {Richter},\ and\ \citenamefont {Rost}}]{TRR00}%
  \BibitemOpen
  \bibfield  {author} {\bibinfo {author} {\bibfnamefont {G.}~\bibnamefont
  {Tanner}}, \bibinfo {author} {\bibfnamefont {K.}~\bibnamefont {Richter}}, \
  and\ \bibinfo {author} {\bibfnamefont {J.~M.}\ \bibnamefont {Rost}},\
  }\href@noop {} {\bibfield  {journal} {\bibinfo  {journal} {Review of Modern
  Physics}\ }\textbf {\bibinfo {volume} {72}},\ \bibinfo {pages} {497}
  (\bibinfo {year} {2000})}\BibitemShut {NoStop}%
\bibitem [{\citenamefont {Haq}\ \emph {et~al.}(1982)\citenamefont {Haq},
  \citenamefont {Pandey},\ and\ \citenamefont {Bohigas}}]{Haq1982}%
  \BibitemOpen
  \bibfield  {author} {\bibinfo {author} {\bibfnamefont {R.~U.}\ \bibnamefont
  {Haq}}, \bibinfo {author} {\bibfnamefont {A.}~\bibnamefont {Pandey}}, \ and\
  \bibinfo {author} {\bibfnamefont {O.}~\bibnamefont {Bohigas}},\ }\href
  {\doibase 10.1103/PhysRevLett.48.1086} {\bibfield  {journal} {\bibinfo
  {journal} {Physical Review Letters}\ }\textbf {\bibinfo {volume} {48}},\
  \bibinfo {pages} {1086} (\bibinfo {year} {1982})}\BibitemShut {NoStop}%
\bibitem [{\citenamefont {Alhassid}(2000)}]{Alhassid2001}%
  \BibitemOpen
  \bibfield  {author} {\bibinfo {author} {\bibfnamefont {Y.}~\bibnamefont
  {Alhassid}},\ }\href {\doibase 10.1103/RevModPhys.72.895} {\bibfield
  {journal} {\bibinfo  {journal} {Reviews of Modern Physics}\ }\textbf
  {\bibinfo {volume} {72}},\ \bibinfo {pages} {895} (\bibinfo {year}
  {2000})}\BibitemShut {NoStop}%
\bibitem [{\citenamefont {Parmantier}(2004)}]{Parmantier2004}%
  \BibitemOpen
  \bibfield  {author} {\bibinfo {author} {\bibfnamefont {J.~P.}\ \bibnamefont
  {Parmantier}},\ }\href {\doibase 10.1109/TEMC.2004.831818} {\bibfield
  {journal} {\bibinfo  {journal} {IEEE Transactions on Electromagnetic
  Compatibility}\ }\textbf {\bibinfo {volume} {46}},\ \bibinfo {pages} {359}
  (\bibinfo {year} {2004})}\BibitemShut {NoStop}%
\bibitem [{\citenamefont {Baum}\ \emph {et~al.}(1978)\citenamefont {Baum},
  \citenamefont {Liu},\ and\ \citenamefont {Tesche}}]{baum1978analysis}%
  \BibitemOpen
  \bibfield  {author} {\bibinfo {author} {\bibfnamefont {C.~E.}\ \bibnamefont
  {Baum}}, \bibinfo {author} {\bibfnamefont {T.~K.}\ \bibnamefont {Liu}}, \
  and\ \bibinfo {author} {\bibfnamefont {F.~M.}\ \bibnamefont {Tesche}},\
  }\href {http://ece-research.unm.edu/summa/notes/In/0350.pdf
  http://ace.unm.edu/summa/notes/In/0350.pdf} {\bibfield  {journal} {\bibinfo
  {journal} {Interaction Note}\ }\textbf {\bibinfo {volume} {350}},\ \bibinfo
  {pages} {467} (\bibinfo {year} {1978})}\BibitemShut {NoStop}%
\bibitem [{\citenamefont {Hill}\ \emph {et~al.}(1994)\citenamefont {Hill},
  \citenamefont {Ma}, \citenamefont {Ondrejka}, \citenamefont {Riddle},
  \citenamefont {Crawford},\ and\ \citenamefont {Johnk}}]{Hill1994}%
  \BibitemOpen
  \bibfield  {author} {\bibinfo {author} {\bibfnamefont {D.}~\bibnamefont
  {Hill}}, \bibinfo {author} {\bibfnamefont {M.}~\bibnamefont {Ma}}, \bibinfo
  {author} {\bibfnamefont {A.}~\bibnamefont {Ondrejka}}, \bibinfo {author}
  {\bibfnamefont {B.}~\bibnamefont {Riddle}}, \bibinfo {author} {\bibfnamefont
  {M.}~\bibnamefont {Crawford}}, \ and\ \bibinfo {author} {\bibfnamefont
  {R.}~\bibnamefont {Johnk}},\ }\href {\doibase 10.1109/15.305461} {\bibfield
  {journal} {\bibinfo  {journal} {IEEE Transactions on Electromagnetic
  Compatibility}\ }\textbf {\bibinfo {volume} {36}},\ \bibinfo {pages} {169}
  (\bibinfo {year} {1994})}\BibitemShut {NoStop}%
\bibitem [{\citenamefont {Hill}(1998)}]{Hill1998}%
  \BibitemOpen
  \bibfield  {author} {\bibinfo {author} {\bibfnamefont {D.}~\bibnamefont
  {Hill}},\ }\href {\doibase 10.1109/15.709418} {\bibfield  {journal} {\bibinfo
   {journal} {IEEE Transactions on Electromagnetic Compatibility}\ }\textbf
  {\bibinfo {volume} {40}},\ \bibinfo {pages} {209} (\bibinfo {year}
  {1998})}\BibitemShut {NoStop}%
\bibitem [{\citenamefont {Junqua}\ \emph {et~al.}(2005)\citenamefont {Junqua},
  \citenamefont {Parmantier},\ and\ \citenamefont {Issac}}]{Junqua2005}%
  \BibitemOpen
  \bibfield  {author} {\bibinfo {author} {\bibfnamefont {I.}~\bibnamefont
  {Junqua}}, \bibinfo {author} {\bibfnamefont {J.-P.}\ \bibnamefont
  {Parmantier}}, \ and\ \bibinfo {author} {\bibfnamefont {F.}~\bibnamefont
  {Issac}},\ }\href {\doibase 10.1080/02726340500214845} {\bibfield  {journal}
  {\bibinfo  {journal} {Electromagnetics}\ }\textbf {\bibinfo {volume} {25}},\
  \bibinfo {pages} {603} (\bibinfo {year} {2005})}\BibitemShut {NoStop}%
\bibitem [{\citenamefont {Tait}\ \emph
  {et~al.}(2011{\natexlab{b}})\citenamefont {Tait}, \citenamefont {Richardson},
  \citenamefont {Slocum}, \citenamefont {Hatfield},\ and\ \citenamefont
  {Rodriguez}}]{Tait2011a}%
  \BibitemOpen
  \bibfield  {author} {\bibinfo {author} {\bibfnamefont {G.~B.}\ \bibnamefont
  {Tait}}, \bibinfo {author} {\bibfnamefont {R.~E.}\ \bibnamefont
  {Richardson}}, \bibinfo {author} {\bibfnamefont {M.~B.}\ \bibnamefont
  {Slocum}}, \bibinfo {author} {\bibfnamefont {M.~O.}\ \bibnamefont
  {Hatfield}}, \ and\ \bibinfo {author} {\bibfnamefont {M.~J.}\ \bibnamefont
  {Rodriguez}},\ }\href {\doibase 10.1109/TEMC.2010.2051442} {\bibfield
  {journal} {\bibinfo  {journal} {IEEE Transactions on Electromagnetic
  Compatibility}\ }\textbf {\bibinfo {volume} {53}},\ \bibinfo {pages} {229}
  (\bibinfo {year} {2011}{\natexlab{b}})}\BibitemShut {NoStop}%
\bibitem [{\citenamefont {Kovalevsky}\ \emph {et~al.}(2015)\citenamefont
  {Kovalevsky}, \citenamefont {Langley}, \citenamefont {Besnier},\ and\
  \citenamefont {Sol}}]{Kovalevsky2015}%
  \BibitemOpen
  \bibfield  {author} {\bibinfo {author} {\bibfnamefont {L.}~\bibnamefont
  {Kovalevsky}}, \bibinfo {author} {\bibfnamefont {R.~S.}\ \bibnamefont
  {Langley}}, \bibinfo {author} {\bibfnamefont {P.}~\bibnamefont {Besnier}}, \
  and\ \bibinfo {author} {\bibfnamefont {J.}~\bibnamefont {Sol}},\ }in\ \href
  {\doibase 10.1109/ISEMC.2015.7256221} {\emph {\bibinfo {booktitle} {2015 IEEE
  International Symposium on Electromagnetic Compatibility (EMC)}}}\ (\bibinfo
  {publisher} {IEEE},\ \bibinfo {year} {2015})\ pp.\ \bibinfo {pages}
  {546--551}\BibitemShut {NoStop}%
\bibitem [{\citenamefont {McKown}\ and\ \citenamefont
  {Hamilton}(1991)}]{McKown1991}%
  \BibitemOpen
  \bibfield  {author} {\bibinfo {author} {\bibfnamefont {J.~W.}\ \bibnamefont
  {McKown}}\ and\ \bibinfo {author} {\bibfnamefont {R.~L.}\ \bibnamefont
  {Hamilton}},\ }\href@noop {} {\bibfield  {journal} {\bibinfo  {journal} {IEEE
  Network Magazine}\ }\textbf {\bibinfo {volume} {27}} (\bibinfo {year}
  {1991})}\BibitemShut {NoStop}%
\bibitem [{\citenamefont {Savioja}\ and\ \citenamefont
  {Svensson}(2015)}]{Savioja2015}%
  \BibitemOpen
  \bibfield  {author} {\bibinfo {author} {\bibfnamefont {L.}~\bibnamefont
  {Savioja}}\ and\ \bibinfo {author} {\bibfnamefont {U.~P.}\ \bibnamefont
  {Svensson}},\ }\href {\doibase 10.1121/1.4926438} {\bibfield  {journal}
  {\bibinfo  {journal} {The Journal of the Acoustical Society of America}\
  }\textbf {\bibinfo {volume} {138}},\ \bibinfo {pages} {708} (\bibinfo {year}
  {2015})}\BibitemShut {NoStop}%
\bibitem [{\citenamefont {Bajars}\ \emph
  {et~al.}(2017{\natexlab{a}})\citenamefont {Bajars}, \citenamefont {Chappell},
  \citenamefont {S{\o}ndergaard},\ and\ \citenamefont {Tanner}}]{Bajars2017}%
  \BibitemOpen
  \bibfield  {author} {\bibinfo {author} {\bibfnamefont {J.}~\bibnamefont
  {Bajars}}, \bibinfo {author} {\bibfnamefont {D.~J.}\ \bibnamefont
  {Chappell}}, \bibinfo {author} {\bibfnamefont {N.}~\bibnamefont
  {S{\o}ndergaard}}, \ and\ \bibinfo {author} {\bibfnamefont {G.}~\bibnamefont
  {Tanner}},\ }\href {\doibase 10.1016/J.JCP.2016.10.019} {\bibfield  {journal}
  {\bibinfo  {journal} {Journal of Computational Physics}\ }\textbf {\bibinfo
  {volume} {328}},\ \bibinfo {pages} {95} (\bibinfo {year}
  {2017}{\natexlab{a}})}\BibitemShut {NoStop}%
\bibitem [{\citenamefont {Bajars}\ \emph
  {et~al.}(2017{\natexlab{b}})\citenamefont {Bajars}, \citenamefont {Chappell},
  \citenamefont {Hartmann},\ and\ \citenamefont {Tanner}}]{Bajars2017a}%
  \BibitemOpen
  \bibfield  {author} {\bibinfo {author} {\bibfnamefont {J.}~\bibnamefont
  {Bajars}}, \bibinfo {author} {\bibfnamefont {D.~J.}\ \bibnamefont
  {Chappell}}, \bibinfo {author} {\bibfnamefont {T.}~\bibnamefont {Hartmann}},
  \ and\ \bibinfo {author} {\bibfnamefont {G.}~\bibnamefont {Tanner}},\ }\href
  {\doibase 10.1007/s10915-017-0397-8} {\bibfield  {journal} {\bibinfo
  {journal} {Journal of Scientific Computing}\ }\textbf {\bibinfo {volume}
  {72}},\ \bibinfo {pages} {1290} (\bibinfo {year}
  {2017}{\natexlab{b}})}\BibitemShut {NoStop}%
\bibitem [{\citenamefont {Hartmann}\ \emph {et~al.}(2019)\citenamefont
  {Hartmann}, \citenamefont {Morita}, \citenamefont {Tanner},\ and\
  \citenamefont {Chappell}}]{HMTC19}%
  \BibitemOpen
  \bibfield  {author} {\bibinfo {author} {\bibfnamefont {T.}~\bibnamefont
  {Hartmann}}, \bibinfo {author} {\bibfnamefont {S.}~\bibnamefont {Morita}},
  \bibinfo {author} {\bibfnamefont {G.}~\bibnamefont {Tanner}}, \ and\ \bibinfo
  {author} {\bibfnamefont {D.~J.}\ \bibnamefont {Chappell}},\ }\href@noop {}
  {\bibfield  {journal} {\bibinfo  {journal} {Wave Motion}\ } (\bibinfo {year}
  {2019})}\BibitemShut {NoStop}%
\bibitem [{\citenamefont {Zheng}\ \emph
  {et~al.}(2006{\natexlab{c}})\citenamefont {Zheng}, \citenamefont {Hemmady},
  \citenamefont {Antonsen}, \citenamefont {Anlage},\ and\ \citenamefont
  {Ott}}]{Zheng2006b}%
  \BibitemOpen
  \bibfield  {author} {\bibinfo {author} {\bibfnamefont {X.}~\bibnamefont
  {Zheng}}, \bibinfo {author} {\bibfnamefont {S.}~\bibnamefont {Hemmady}},
  \bibinfo {author} {\bibfnamefont {T.~M.}\ \bibnamefont {Antonsen}}, \bibinfo
  {author} {\bibfnamefont {S.~M.}\ \bibnamefont {Anlage}}, \ and\ \bibinfo
  {author} {\bibfnamefont {E.}~\bibnamefont {Ott}},\ }\href {\doibase
  10.1103/PhysRevE.73.046208} {\bibfield  {journal} {\bibinfo  {journal}
  {Physical Review E}\ }\textbf {\bibinfo {volume} {73}},\ \bibinfo {pages}
  {046208} (\bibinfo {year} {2006}{\natexlab{c}})}\BibitemShut {NoStop}%
\bibitem [{\citenamefont {Casati}\ \emph {et~al.}(1980)\citenamefont {Casati},
  \citenamefont {Valz-Gris},\ and\ \citenamefont {Guarnieri}}]{Casati1980}%
  \BibitemOpen
  \bibfield  {author} {\bibinfo {author} {\bibfnamefont {G.}~\bibnamefont
  {Casati}}, \bibinfo {author} {\bibfnamefont {F.}~\bibnamefont {Valz-Gris}}, \
  and\ \bibinfo {author} {\bibfnamefont {I.}~\bibnamefont {Guarnieri}},\ }\href
  {\doibase 10.1007/BF02798790} {\bibfield  {journal} {\bibinfo  {journal}
  {Lettere al Nuovo Cimento}\ }\textbf {\bibinfo {volume} {28}},\ \bibinfo
  {pages} {279} (\bibinfo {year} {1980})}\BibitemShut {NoStop}%
\bibitem [{\citenamefont {Bohigas}\ \emph {et~al.}(1984)\citenamefont
  {Bohigas}, \citenamefont {Giannoni},\ and\ \citenamefont
  {Schmit}}]{Bohigas1984}%
  \BibitemOpen
  \bibfield  {author} {\bibinfo {author} {\bibfnamefont {O.}~\bibnamefont
  {Bohigas}}, \bibinfo {author} {\bibfnamefont {M.~J.}\ \bibnamefont
  {Giannoni}}, \ and\ \bibinfo {author} {\bibfnamefont {C.}~\bibnamefont
  {Schmit}},\ }\href {\doibase 10.1103/PhysRevLett.52.1} {\bibfield  {journal}
  {\bibinfo  {journal} {Physical Review Letters}\ }\textbf {\bibinfo {volume}
  {52}},\ \bibinfo {pages} {1} (\bibinfo {year} {1984})}\BibitemShut {NoStop}%
\bibitem [{\citenamefont {Hart}\ \emph {et~al.}(2009)\citenamefont {Hart},
  \citenamefont {Antonsen},\ and\ \citenamefont {Ott}}]{Hart2009}%
  \BibitemOpen
  \bibfield  {author} {\bibinfo {author} {\bibfnamefont {J.~A.}\ \bibnamefont
  {Hart}}, \bibinfo {author} {\bibfnamefont {T.~M.}\ \bibnamefont {Antonsen}},
  \ and\ \bibinfo {author} {\bibfnamefont {E.}~\bibnamefont {Ott}},\ }\href
  {\doibase 10.1103/PhysRevE.80.041109} {\bibfield  {journal} {\bibinfo
  {journal} {Physical Review E}\ }\textbf {\bibinfo {volume} {80}},\ \bibinfo
  {pages} {041109} (\bibinfo {year} {2009})}\BibitemShut {NoStop}%
\bibitem [{\citenamefont {Yeh}\ \emph {et~al.}(2010)\citenamefont {Yeh},
  \citenamefont {Hart}, \citenamefont {Bradshaw}, \citenamefont {Antonsen},
  \citenamefont {Ott},\ and\ \citenamefont {Anlage}}]{Yeh}%
  \BibitemOpen
  \bibfield  {author} {\bibinfo {author} {\bibfnamefont {J.-H.}\ \bibnamefont
  {Yeh}}, \bibinfo {author} {\bibfnamefont {J.~A.}\ \bibnamefont {Hart}},
  \bibinfo {author} {\bibfnamefont {E.}~\bibnamefont {Bradshaw}}, \bibinfo
  {author} {\bibfnamefont {T.~M.}\ \bibnamefont {Antonsen}}, \bibinfo {author}
  {\bibfnamefont {E.}~\bibnamefont {Ott}}, \ and\ \bibinfo {author}
  {\bibfnamefont {S.~M.}\ \bibnamefont {Anlage}},\ }\href {\doibase
  10.1103/PhysRevE.82.041114} {\bibfield  {journal} {\bibinfo  {journal}
  {Physical Review E}\ }\textbf {\bibinfo {volume} {82}},\ \bibinfo {pages}
  {041114} (\bibinfo {year} {2010})}\BibitemShut {NoStop}%
\bibitem [{\citenamefont {Gradoni}\ \emph {et~al.}(2014)\citenamefont
  {Gradoni}, \citenamefont {Yeh}, \citenamefont {Xiao}, \citenamefont
  {Antonsen}, \citenamefont {Anlage},\ and\ \citenamefont {Ott}}]{Gradoni2014}%
  \BibitemOpen
  \bibfield  {author} {\bibinfo {author} {\bibfnamefont {G.}~\bibnamefont
  {Gradoni}}, \bibinfo {author} {\bibfnamefont {J.-H.}\ \bibnamefont {Yeh}},
  \bibinfo {author} {\bibfnamefont {B.}~\bibnamefont {Xiao}}, \bibinfo {author}
  {\bibfnamefont {T.~M.}\ \bibnamefont {Antonsen}}, \bibinfo {author}
  {\bibfnamefont {S.~M.}\ \bibnamefont {Anlage}}, \ and\ \bibinfo {author}
  {\bibfnamefont {E.}~\bibnamefont {Ott}},\ }\href {\doibase
  10.1016/j.wavemoti.2014.02.003} {\bibfield  {journal} {\bibinfo  {journal}
  {Wave Motion}\ }\textbf {\bibinfo {volume} {51}},\ \bibinfo {pages} {606}
  (\bibinfo {year} {2014})}\BibitemShut {NoStop}%
\bibitem [{\citenamefont {Xiao}\ \emph {et~al.}(2016)\citenamefont {Xiao},
  \citenamefont {Antonsen}, \citenamefont {Ott},\ and\ \citenamefont
  {Anlage}}]{Xiao2016}%
  \BibitemOpen
  \bibfield  {author} {\bibinfo {author} {\bibfnamefont {B.}~\bibnamefont
  {Xiao}}, \bibinfo {author} {\bibfnamefont {T.~M.}\ \bibnamefont {Antonsen}},
  \bibinfo {author} {\bibfnamefont {E.}~\bibnamefont {Ott}}, \ and\ \bibinfo
  {author} {\bibfnamefont {S.~M.}\ \bibnamefont {Anlage}},\ }\href {\doibase
  10.1103/PhysRevE.93.052205} {\bibfield  {journal} {\bibinfo  {journal}
  {Physical Review E}\ }\textbf {\bibinfo {volume} {93}},\ \bibinfo {pages}
  {052205} (\bibinfo {year} {2016})}\BibitemShut {NoStop}%
\bibitem [{\citenamefont {Ma}\ \emph {et~al.}(2020)\citenamefont {Ma},
  \citenamefont {Xiao}, \citenamefont {Drikas}, \citenamefont {Addissie},
  \citenamefont {Hong}, \citenamefont {Antonsen}, \citenamefont {Ott},\ and\
  \citenamefont {Anlage}}]{ma2020}%
  \BibitemOpen
  \bibfield  {author} {\bibinfo {author} {\bibfnamefont {S.}~\bibnamefont
  {Ma}}, \bibinfo {author} {\bibfnamefont {B.}~\bibnamefont {Xiao}}, \bibinfo
  {author} {\bibfnamefont {Z.}~\bibnamefont {Drikas}}, \bibinfo {author}
  {\bibfnamefont {B.}~\bibnamefont {Addissie}}, \bibinfo {author}
  {\bibfnamefont {R.}~\bibnamefont {Hong}}, \bibinfo {author} {\bibfnamefont
  {T.~M.}\ \bibnamefont {Antonsen}}, \bibinfo {author} {\bibfnamefont
  {E.}~\bibnamefont {Ott}}, \ and\ \bibinfo {author} {\bibfnamefont {S.~M.}\
  \bibnamefont {Anlage}},\ }\href {\doibase 10.1103/PhysRevE.101.022201}
  {\bibfield  {journal} {\bibinfo  {journal} {Physical Review E}\ }\textbf
  {\bibinfo {volume} {101}},\ \bibinfo {pages} {022201} (\bibinfo {year}
  {2020})}\BibitemShut {NoStop}%
\bibitem [{\citenamefont {Zhou}\ \emph {et~al.}(2017)\citenamefont {Zhou},
  \citenamefont {Ott}, \citenamefont {Antonsen},\ and\ \citenamefont
  {Anlage}}]{Zhou2017}%
  \BibitemOpen
  \bibfield  {author} {\bibinfo {author} {\bibfnamefont {M.}~\bibnamefont
  {Zhou}}, \bibinfo {author} {\bibfnamefont {E.}~\bibnamefont {Ott}}, \bibinfo
  {author} {\bibfnamefont {T.~M.}\ \bibnamefont {Antonsen}}, \ and\ \bibinfo
  {author} {\bibfnamefont {S.~M.}\ \bibnamefont {Anlage}},\ }\href {\doibase
  10.1063/1.4986499} {\bibfield  {journal} {\bibinfo  {journal} {Chaos}\
  }\textbf {\bibinfo {volume} {27}},\ \bibinfo {pages} {103114} (\bibinfo
  {year} {2017})}\BibitemShut {NoStop}%
\bibitem [{\citenamefont {Zhou}\ \emph {et~al.}(2019)\citenamefont {Zhou},
  \citenamefont {Ott}, \citenamefont {Antonsen},\ and\ \citenamefont
  {Anlage}}]{Zhou2019}%
  \BibitemOpen
  \bibfield  {author} {\bibinfo {author} {\bibfnamefont {M.}~\bibnamefont
  {Zhou}}, \bibinfo {author} {\bibfnamefont {E.}~\bibnamefont {Ott}}, \bibinfo
  {author} {\bibfnamefont {T.~M.}\ \bibnamefont {Antonsen}}, \ and\ \bibinfo
  {author} {\bibfnamefont {S.~M.}\ \bibnamefont {Anlage}},\ }\href {\doibase
  10.1063/1.5085653} {\bibfield  {journal} {\bibinfo  {journal} {Chaos: An
  Interdisciplinary Journal of Nonlinear Science}\ }\textbf {\bibinfo {volume}
  {29}},\ \bibinfo {pages} {033113} (\bibinfo {year} {2019})}\BibitemShut
  {NoStop}%
\bibitem [{\citenamefont {Gradoni}\ \emph {et~al.}(2015)\citenamefont
  {Gradoni}, \citenamefont {Antonsen}, \citenamefont {Anlage},\ and\
  \citenamefont {Ott}}]{Gradoni2015}%
  \BibitemOpen
  \bibfield  {author} {\bibinfo {author} {\bibfnamefont {G.}~\bibnamefont
  {Gradoni}}, \bibinfo {author} {\bibfnamefont {T.~M.}\ \bibnamefont
  {Antonsen}}, \bibinfo {author} {\bibfnamefont {S.~M.}\ \bibnamefont
  {Anlage}}, \ and\ \bibinfo {author} {\bibfnamefont {E.}~\bibnamefont {Ott}},\
  }\href {\doibase 10.1109/TEMC.2015.2421346} {\bibfield  {journal} {\bibinfo
  {journal} {IEEE Transactions on Electromagnetic Compatibility}\ }\textbf
  {\bibinfo {volume} {57}},\ \bibinfo {pages} {1049} (\bibinfo {year}
  {2015})}\BibitemShut {NoStop}%
\bibitem [{\citenamefont {Hoijer}\ and\ \citenamefont
  {Kroon}(2013)}]{Hoijer2013}%
  \BibitemOpen
  \bibfield  {author} {\bibinfo {author} {\bibfnamefont {M.}~\bibnamefont
  {Hoijer}}\ and\ \bibinfo {author} {\bibfnamefont {L.}~\bibnamefont {Kroon}},\
  }\href {\doibase 10.1109/TEMC.2013.2249510} {\bibfield  {journal} {\bibinfo
  {journal} {IEEE Transactions on Electromagnetic Compatibility}\ }\textbf
  {\bibinfo {volume} {55}},\ \bibinfo {pages} {1328} (\bibinfo {year}
  {2013})}\BibitemShut {NoStop}%
\bibitem [{\citenamefont {Serra}\ \emph {et~al.}(2017)\citenamefont {Serra},
  \citenamefont {Marvin}, \citenamefont {Moglie}, \citenamefont {Primiani},
  \citenamefont {Cozza}, \citenamefont {Arnaut}, \citenamefont {Huang},
  \citenamefont {Hatfield}, \citenamefont {Klingler},\ and\ \citenamefont
  {Leferink}}]{Serra2017}%
  \BibitemOpen
  \bibfield  {author} {\bibinfo {author} {\bibfnamefont {R.}~\bibnamefont
  {Serra}}, \bibinfo {author} {\bibfnamefont {A.~C.}\ \bibnamefont {Marvin}},
  \bibinfo {author} {\bibfnamefont {F.}~\bibnamefont {Moglie}}, \bibinfo
  {author} {\bibfnamefont {V.~M.}\ \bibnamefont {Primiani}}, \bibinfo {author}
  {\bibfnamefont {A.}~\bibnamefont {Cozza}}, \bibinfo {author} {\bibfnamefont
  {L.~R.}\ \bibnamefont {Arnaut}}, \bibinfo {author} {\bibfnamefont
  {Y.}~\bibnamefont {Huang}}, \bibinfo {author} {\bibfnamefont {M.~O.}\
  \bibnamefont {Hatfield}}, \bibinfo {author} {\bibfnamefont {M.}~\bibnamefont
  {Klingler}}, \ and\ \bibinfo {author} {\bibfnamefont {F.}~\bibnamefont
  {Leferink}},\ }\href {\doibase 10.1109/MEMC.2017.7931986} {\bibfield
  {journal} {\bibinfo  {journal} {IEEE Electromagnetic Compatibility Magazine}\
  }\textbf {\bibinfo {volume} {6}},\ \bibinfo {pages} {63} (\bibinfo {year}
  {2017})}\BibitemShut {NoStop}%
\bibitem [{\citenamefont {Frazier}\ \emph
  {et~al.}(2013{\natexlab{a}})\citenamefont {Frazier}, \citenamefont {Taddese},
  \citenamefont {Antonsen},\ and\ \citenamefont {Anlage}}]{Frazier2013}%
  \BibitemOpen
  \bibfield  {author} {\bibinfo {author} {\bibfnamefont {M.}~\bibnamefont
  {Frazier}}, \bibinfo {author} {\bibfnamefont {B.}~\bibnamefont {Taddese}},
  \bibinfo {author} {\bibfnamefont {T.}~\bibnamefont {Antonsen}}, \ and\
  \bibinfo {author} {\bibfnamefont {S.~M.}\ \bibnamefont {Anlage}},\ }\href
  {\doibase 10.1103/PhysRevLett.110.063902} {\bibfield  {journal} {\bibinfo
  {journal} {Physical Review Letters}\ }\textbf {\bibinfo {volume} {110}},\
  \bibinfo {pages} {063902} (\bibinfo {year} {2013}{\natexlab{a}})}\BibitemShut
  {NoStop}%
\bibitem [{\citenamefont {Frazier}\ \emph
  {et~al.}(2013{\natexlab{b}})\citenamefont {Frazier}, \citenamefont {Taddese},
  \citenamefont {Xiao}, \citenamefont {Antonsen}, \citenamefont {Ott},\ and\
  \citenamefont {Anlage}}]{Frazier2013a}%
  \BibitemOpen
  \bibfield  {author} {\bibinfo {author} {\bibfnamefont {M.}~\bibnamefont
  {Frazier}}, \bibinfo {author} {\bibfnamefont {B.}~\bibnamefont {Taddese}},
  \bibinfo {author} {\bibfnamefont {B.}~\bibnamefont {Xiao}}, \bibinfo {author}
  {\bibfnamefont {T.}~\bibnamefont {Antonsen}}, \bibinfo {author}
  {\bibfnamefont {E.}~\bibnamefont {Ott}}, \ and\ \bibinfo {author}
  {\bibfnamefont {S.~M.}\ \bibnamefont {Anlage}},\ }\href {\doibase
  10.1103/PhysRevE.88.062910} {\bibfield  {journal} {\bibinfo  {journal}
  {Physical Review E}\ }\textbf {\bibinfo {volume} {88}},\ \bibinfo {pages}
  {062910} (\bibinfo {year} {2013}{\natexlab{b}})}\BibitemShut {NoStop}%
\bibitem [{\citenamefont {Hemmady}\ \emph {et~al.}(2006)\citenamefont
  {Hemmady}, \citenamefont {Zheng}, \citenamefont {Hart}, \citenamefont
  {Antonsen}, \citenamefont {Ott},\ and\ \citenamefont {Anlage}}]{Hemmady}%
  \BibitemOpen
  \bibfield  {author} {\bibinfo {author} {\bibfnamefont {S.}~\bibnamefont
  {Hemmady}}, \bibinfo {author} {\bibfnamefont {X.}~\bibnamefont {Zheng}},
  \bibinfo {author} {\bibfnamefont {J.}~\bibnamefont {Hart}}, \bibinfo {author}
  {\bibfnamefont {T.~M.}\ \bibnamefont {Antonsen}}, \bibinfo {author}
  {\bibfnamefont {E.}~\bibnamefont {Ott}}, \ and\ \bibinfo {author}
  {\bibfnamefont {S.~M.}\ \bibnamefont {Anlage}},\ }\href {\doibase
  10.1103/PhysRevE.74.036213} {\bibfield  {journal} {\bibinfo  {journal}
  {Physical Review E}\ }\textbf {\bibinfo {volume} {74}},\ \bibinfo {pages}
  {036213} (\bibinfo {year} {2006})}\BibitemShut {NoStop}%
\bibitem [{\citenamefont {Drikas}\ \emph {et~al.}(2014)\citenamefont {Drikas},
  \citenamefont {{Gil Gil}}, \citenamefont {Hong}, \citenamefont {Andreadis},
  \citenamefont {Yeh}, \citenamefont {Taddese},\ and\ \citenamefont
  {Anlage}}]{Drikas2014}%
  \BibitemOpen
  \bibfield  {author} {\bibinfo {author} {\bibfnamefont {Z.~B.}\ \bibnamefont
  {Drikas}}, \bibinfo {author} {\bibfnamefont {J.}~\bibnamefont {{Gil Gil}}},
  \bibinfo {author} {\bibfnamefont {S.~K.}\ \bibnamefont {Hong}}, \bibinfo
  {author} {\bibfnamefont {T.~D.}\ \bibnamefont {Andreadis}}, \bibinfo {author}
  {\bibfnamefont {J.-H.}\ \bibnamefont {Yeh}}, \bibinfo {author} {\bibfnamefont
  {B.~T.}\ \bibnamefont {Taddese}}, \ and\ \bibinfo {author} {\bibfnamefont
  {S.~M.}\ \bibnamefont {Anlage}},\ }\href {\doibase 10.1109/TEMC.2014.2337262}
  {\bibfield  {journal} {\bibinfo  {journal} {IEEE Transactions on
  Electromagnetic Compatibility}\ }\textbf {\bibinfo {volume} {56}},\ \bibinfo
  {pages} {1480} (\bibinfo {year} {2014})}\BibitemShut {NoStop}%
\bibitem [{\citenamefont {Berry}(1981)}]{Berry1981}%
  \BibitemOpen
  \bibfield  {author} {\bibinfo {author} {\bibfnamefont {M.~V.}\ \bibnamefont
  {Berry}},\ }\href {\doibase 10.1088/0143-0807/2/2/006} {\bibfield  {journal}
  {\bibinfo  {journal} {European Journal of Physics}\ }\textbf {\bibinfo
  {volume} {2}},\ \bibinfo {pages} {91} (\bibinfo {year} {1981})}\BibitemShut
  {NoStop}%
\bibitem [{\citenamefont {Fyodorov}\ and\ \citenamefont
  {Savin}(2004)}]{Fyodorov2004}%
  \BibitemOpen
  \bibfield  {author} {\bibinfo {author} {\bibfnamefont {Y.~V.}\ \bibnamefont
  {Fyodorov}}\ and\ \bibinfo {author} {\bibfnamefont {D.~V.}\ \bibnamefont
  {Savin}},\ }\href {\doibase 10.1134/1.1868794} {\bibfield  {journal}
  {\bibinfo  {journal} {Journal of Experimental and Theoretical Physics
  Letters}\ }\textbf {\bibinfo {volume} {80}},\ \bibinfo {pages} {725}
  (\bibinfo {year} {2004})}\BibitemShut {NoStop}%
\bibitem [{\citenamefont {Rehemanjiang}\ \emph {et~al.}(2016)\citenamefont
  {Rehemanjiang}, \citenamefont {Allgaier}, \citenamefont {Joyner},
  \citenamefont {M{\"{u}}ller}, \citenamefont {Sieber}, \citenamefont {Kuhl},\
  and\ \citenamefont {St{\"{o}}ckmann}}]{Rehemanjiang2016}%
  \BibitemOpen
  \bibfield  {author} {\bibinfo {author} {\bibfnamefont {A.}~\bibnamefont
  {Rehemanjiang}}, \bibinfo {author} {\bibfnamefont {M.}~\bibnamefont
  {Allgaier}}, \bibinfo {author} {\bibfnamefont {C.~H.}\ \bibnamefont
  {Joyner}}, \bibinfo {author} {\bibfnamefont {S.}~\bibnamefont
  {M{\"{u}}ller}}, \bibinfo {author} {\bibfnamefont {M.}~\bibnamefont
  {Sieber}}, \bibinfo {author} {\bibfnamefont {U.}~\bibnamefont {Kuhl}}, \ and\
  \bibinfo {author} {\bibfnamefont {H.-J.}\ \bibnamefont {St{\"{o}}ckmann}},\
  }\href {\doibase 10.1103/PhysRevLett.117.064101} {\bibfield  {journal}
  {\bibinfo  {journal} {Physical Review Letters}\ }\textbf {\bibinfo {volume}
  {117}},\ \bibinfo {pages} {064101} (\bibinfo {year} {2016})}\BibitemShut
  {NoStop}%
\bibitem [{\citenamefont {Coppersmith}\ and\ \citenamefont
  {Winograd}(1990)}]{Coppersmith1990}%
  \BibitemOpen
  \bibfield  {author} {\bibinfo {author} {\bibfnamefont {D.}~\bibnamefont
  {Coppersmith}}\ and\ \bibinfo {author} {\bibfnamefont {S.}~\bibnamefont
  {Winograd}},\ }\href {\doibase 10.1016/S0747-7171(08)80013-2} {\bibfield
  {journal} {\bibinfo  {journal} {Journal of Symbolic Computation}\ }\textbf
  {\bibinfo {volume} {9}},\ \bibinfo {pages} {251} (\bibinfo {year}
  {1990})}\BibitemShut {NoStop}%
\bibitem [{\citenamefont {Gunnarsson}\ and\ \citenamefont
  {Backstrom}(2014)}]{Gunnarsson2014}%
  \BibitemOpen
  \bibfield  {author} {\bibinfo {author} {\bibfnamefont {R.}~\bibnamefont
  {Gunnarsson}}\ and\ \bibinfo {author} {\bibfnamefont {M.}~\bibnamefont
  {Backstrom}},\ }in\ \href {\doibase 10.1109/EMCEurope.2014.6930897} {\emph
  {\bibinfo {booktitle} {2014 International Symposium on Electromagnetic
  Compatibility}}}\ (\bibinfo  {publisher} {IEEE},\ \bibinfo {year} {2014})\
  pp.\ \bibinfo {pages} {169--174}\BibitemShut {NoStop}%
\bibitem [{\citenamefont {{Gil Gil}}\ \emph {et~al.}(2016)\citenamefont {{Gil
  Gil}}, \citenamefont {Drikas}, \citenamefont {Andreadis},\ and\ \citenamefont
  {Anlage}}]{GilGil2016}%
  \BibitemOpen
  \bibfield  {author} {\bibinfo {author} {\bibfnamefont {J.}~\bibnamefont {{Gil
  Gil}}}, \bibinfo {author} {\bibfnamefont {Z.~B.}\ \bibnamefont {Drikas}},
  \bibinfo {author} {\bibfnamefont {T.~D.}\ \bibnamefont {Andreadis}}, \ and\
  \bibinfo {author} {\bibfnamefont {S.~M.}\ \bibnamefont {Anlage}},\ }\href
  {\doibase 10.1109/TEMC.2016.2580301} {\bibfield  {journal} {\bibinfo
  {journal} {IEEE Transactions on Electromagnetic Compatibility}\ }\textbf
  {\bibinfo {volume} {58}},\ \bibinfo {pages} {1535} (\bibinfo {year}
  {2016})}\BibitemShut {NoStop}%
\bibitem [{\citenamefont {Yeh}\ \emph {et~al.}(2012)\citenamefont {Yeh},
  \citenamefont {Antonsen}, \citenamefont {Ott},\ and\ \citenamefont
  {Anlage}}]{Yeh2012}%
  \BibitemOpen
  \bibfield  {author} {\bibinfo {author} {\bibfnamefont {J.-H.}\ \bibnamefont
  {Yeh}}, \bibinfo {author} {\bibfnamefont {T.~M.}\ \bibnamefont {Antonsen}},
  \bibinfo {author} {\bibfnamefont {E.}~\bibnamefont {Ott}}, \ and\ \bibinfo
  {author} {\bibfnamefont {S.~M.}\ \bibnamefont {Anlage}},\ }\href {\doibase
  10.1103/PhysRevE.85.015202} {\bibfield  {journal} {\bibinfo  {journal}
  {Physical Review E}\ }\textbf {\bibinfo {volume} {85}},\ \bibinfo {pages}
  {015202} (\bibinfo {year} {2012})}\BibitemShut {NoStop}%
\bibitem [{\citenamefont {Yeh}\ \emph {et~al.}(2013)\citenamefont {Yeh},
  \citenamefont {Drikas}, \citenamefont {{Gil Gil}}, \citenamefont {Hong},
  \citenamefont {Taddese}, \citenamefont {Ott}, \citenamefont {Antonsen},
  \citenamefont {Andreadis},\ and\ \citenamefont {Anlage}}]{Yeh2013}%
  \BibitemOpen
  \bibfield  {author} {\bibinfo {author} {\bibfnamefont {J.-H.}\ \bibnamefont
  {Yeh}}, \bibinfo {author} {\bibfnamefont {Z.}~\bibnamefont {Drikas}},
  \bibinfo {author} {\bibfnamefont {J.}~\bibnamefont {{Gil Gil}}}, \bibinfo
  {author} {\bibfnamefont {S.}~\bibnamefont {Hong}}, \bibinfo {author}
  {\bibfnamefont {B.}~\bibnamefont {Taddese}}, \bibinfo {author} {\bibfnamefont
  {E.}~\bibnamefont {Ott}}, \bibinfo {author} {\bibfnamefont {T.}~\bibnamefont
  {Antonsen}}, \bibinfo {author} {\bibfnamefont {T.}~\bibnamefont {Andreadis}},
  \ and\ \bibinfo {author} {\bibfnamefont {S.}~\bibnamefont {Anlage}},\ }\href
  {\doibase 10.12693/APhysPolA.124.1045} {\bibfield  {journal} {\bibinfo
  {journal} {Acta Physica Polonica A}\ }\textbf {\bibinfo {volume} {124}},\
  \bibinfo {pages} {1045} (\bibinfo {year} {2013})}\BibitemShut {NoStop}%
\bibitem [{\citenamefont {Yan}\ \emph {et~al.}(2019)\citenamefont {Yan},
  \citenamefont {Dawson}, \citenamefont {Marvin}, \citenamefont {Flintoft},\
  and\ \citenamefont {Robinson}}]{Yan2019}%
  \BibitemOpen
  \bibfield  {author} {\bibinfo {author} {\bibfnamefont {J.}~\bibnamefont
  {Yan}}, \bibinfo {author} {\bibfnamefont {J.~F.}\ \bibnamefont {Dawson}},
  \bibinfo {author} {\bibfnamefont {A.~C.}\ \bibnamefont {Marvin}}, \bibinfo
  {author} {\bibfnamefont {I.~D.}\ \bibnamefont {Flintoft}}, \ and\ \bibinfo
  {author} {\bibfnamefont {M.~P.}\ \bibnamefont {Robinson}},\ }\href {\doibase
  10.1109/TEMC.2019.2916820} {\bibfield  {journal} {\bibinfo  {journal} {IEEE
  Transactions on Electromagnetic Compatibility}\ }\textbf {\bibinfo {volume}
  {61}},\ \bibinfo {pages} {1362} (\bibinfo {year} {2019})}\BibitemShut
  {NoStop}%
\bibitem [{\citenamefont {Wu}\ \emph {et~al.}(1998)\citenamefont {Wu},
  \citenamefont {Bridgewater}, \citenamefont {Gokirmak},\ and\ \citenamefont
  {Anlage}}]{Wu1998}%
  \BibitemOpen
  \bibfield  {author} {\bibinfo {author} {\bibfnamefont {D.~H.}\ \bibnamefont
  {Wu}}, \bibinfo {author} {\bibfnamefont {J.~S.~A.}\ \bibnamefont
  {Bridgewater}}, \bibinfo {author} {\bibfnamefont {A.}~\bibnamefont
  {Gokirmak}}, \ and\ \bibinfo {author} {\bibfnamefont {S.~M.}\ \bibnamefont
  {Anlage}},\ }\href {\doibase 10.1103/PhysRevLett.81.2890} {\bibfield
  {journal} {\bibinfo  {journal} {Physical Review Letters}\ }\textbf {\bibinfo
  {volume} {81}},\ \bibinfo {pages} {2890} (\bibinfo {year}
  {1998})}\BibitemShut {NoStop}%
\bibitem [{\citenamefont {Farasartul}\ \emph {et~al.}(2020)\citenamefont
  {Farasartul}, \citenamefont {Blakaj}, \citenamefont {Phang}, \citenamefont
  {Antonsen}, \citenamefont {Creagh}, \citenamefont {Gradoni},\ and\
  \citenamefont {Tanner}}]{Adnan2020}%
  \BibitemOpen
  \bibfield  {author} {\bibinfo {author} {\bibfnamefont {A.}~\bibnamefont
  {Farasartul}}, \bibinfo {author} {\bibfnamefont {V.}~\bibnamefont {Blakaj}},
  \bibinfo {author} {\bibfnamefont {S.}~\bibnamefont {Phang}}, \bibinfo
  {author} {\bibfnamefont {T.~M.}\ \bibnamefont {Antonsen}}, \bibinfo {author}
  {\bibfnamefont {S.~C.}\ \bibnamefont {Creagh}}, \bibinfo {author}
  {\bibfnamefont {G.}~\bibnamefont {Gradoni}}, \ and\ \bibinfo {author}
  {\bibfnamefont {G.}~\bibnamefont {Tanner}},\ }\href@noop {} {\bibfield
  {journal} {\bibinfo  {journal} {to appear in Proceedings of the Royal Society
  A}\ } (\bibinfo {year} {2020})}\BibitemShut {NoStop}%
\bibitem [{\citenamefont {Yeh}\ and\ \citenamefont {Anlage}(2013)}]{Yeh2013a}%
  \BibitemOpen
  \bibfield  {author} {\bibinfo {author} {\bibfnamefont {J.-H.}\ \bibnamefont
  {Yeh}}\ and\ \bibinfo {author} {\bibfnamefont {S.~M.}\ \bibnamefont
  {Anlage}},\ }\href {\doibase 10.1063/1.4797461} {\bibfield  {journal}
  {\bibinfo  {journal} {Review of Scientific Instruments}\ }\textbf {\bibinfo
  {volume} {84}},\ \bibinfo {pages} {034706} (\bibinfo {year}
  {2013})}\BibitemShut {NoStop}%
\bibitem [{\citenamefont {Lin}\ \emph {et~al.}(2019)\citenamefont {Lin},
  \citenamefont {Peng}, \citenamefont {Schamiloglu}, \citenamefont {Drikas},\
  and\ \citenamefont {Antonsen}}]{Lin2019}%
  \BibitemOpen
  \bibfield  {author} {\bibinfo {author} {\bibfnamefont {S.}~\bibnamefont
  {Lin}}, \bibinfo {author} {\bibfnamefont {Z.}~\bibnamefont {Peng}}, \bibinfo
  {author} {\bibfnamefont {E.}~\bibnamefont {Schamiloglu}}, \bibinfo {author}
  {\bibfnamefont {Z.~B.}\ \bibnamefont {Drikas}}, \ and\ \bibinfo {author}
  {\bibfnamefont {T.}~\bibnamefont {Antonsen}},\ }in\ \href {\doibase
  10.1109/ISEMC.2019.8825194} {\emph {\bibinfo {booktitle} {2019 IEEE
  International Symposium on Electromagnetic Compatibility, Signal {\&} Power
  Integrity (EMC+SIPI)}}}\ (\bibinfo  {publisher} {IEEE},\ \bibinfo {year}
  {2019})\ pp.\ \bibinfo {pages} {499--504}\BibitemShut {NoStop}%
\end{thebibliography}

%

\end{document}